\documentclass[12pt,oneside,,superscriptaddress]{revtex4-2}
\usepackage{lineno} % list the numbers of lines
\usepackage{amssymb}
\usepackage{footmisc}
\usepackage{natbib}
\usepackage{anysize}
\usepackage{graphicx}
\usepackage{amsmath}
\usepackage{wasysym}

\usepackage{setspace}
\usepackage[bookmarks = false]{hyperref}
\usepackage{color}
\usepackage{subfigure}
\usepackage{multirow}
\usepackage[T1]{fontenc}
\renewcommand{\baselinestretch}{1.5}
\renewcommand{\andname}{\ignorespaces}
\marginsize{2.3cm}{2.3cm}{1.5cm}{1.5cm}
\usepackage{color}
\definecolor{SH}{RGB}{0,0,200}
\definecolor{SH2}{RGB}{0,200,200}

\usepackage{setspace}
\makeatletter

\makeatletter

\begin{document}

%less 15 words
\title{Chiral quantum heating and cooling with an optically controlled ion}
\author{J.-T.~Bu}  \thanks{Co-first authors with equal contribution}
\affiliation{State Key Laboratory of Magnetic Resonance and Atomic and Molecular Physics, Wuhan Institute of Physics and Mathematics, Innovation Academy of Precision Measurement Science and Technology, Chinese Academy of Sciences, Wuhan 430071, China}
\affiliation{University of the Chinese Academy of Sciences, Beijing 100049, China}
\author{J.-Q.~Zhang} \thanks{Co-first authors with equal contribution}
\affiliation{State Key Laboratory of Magnetic Resonance and Atomic and Molecular Physics, Wuhan Institute of Physics and Mathematics, Innovation Academy of Precision Measurement Science and Technology, Chinese Academy of Sciences, Wuhan 430071, China}
\author{G.-Y.~Ding}  \thanks{Co-first authors with equal contribution}
\affiliation{State Key Laboratory of Magnetic Resonance and Atomic and Molecular Physics, Wuhan Institute of Physics and Mathematics, Innovation Academy of Precision Measurement Science and Technology, Chinese Academy of Sciences, Wuhan 430071, China}
\affiliation{University of the Chinese Academy of Sciences, Beijing 100049, China}
\author{J.-C.~Li}
\affiliation{State Key Laboratory of Magnetic Resonance and Atomic and Molecular Physics, Wuhan Institute of Physics and Mathematics, Innovation Academy of Precision Measurement Science and Technology, Chinese Academy of Sciences, Wuhan 430071, China}
\affiliation{University of the Chinese Academy of Sciences, Beijing 100049, China}
\author{J.-W.~Zhang}
\affiliation{Research Center for Quantum Precision Measurement, Guangzhou Institute of Industry Technology, Guangzhou, 511458, China }
\author{B.~Wang}
\affiliation{State Key Laboratory of Magnetic Resonance and Atomic and Molecular Physics, Wuhan Institute of Physics and Mathematics, Innovation Academy of Precision Measurement Science and Technology, Chinese Academy of Sciences, Wuhan 430071, China}
\affiliation{University of the Chinese Academy of Sciences, Beijing 100049, China}
\author{W.-Q.~Ding}
\affiliation{State Key Laboratory of Magnetic Resonance and Atomic and Molecular Physics, Wuhan Institute of Physics and Mathematics, Innovation Academy of Precision Measurement Science and Technology, Chinese Academy of Sciences, Wuhan 430071, China}
\affiliation{University of the Chinese Academy of Sciences, Beijing 100049, China}
\author{W.-F.~Yuan}
\affiliation{State Key Laboratory of Magnetic Resonance and Atomic and Molecular Physics, Wuhan Institute of Physics and Mathematics, Innovation Academy of Precision Measurement Science and Technology, Chinese Academy of Sciences, Wuhan 430071, China}
\affiliation{University of the Chinese Academy of Sciences, Beijing 100049, China}
\author{L.~Chen}
\affiliation{State Key Laboratory of Magnetic Resonance and Atomic and Molecular Physics, Wuhan Institute of Physics and Mathematics, Innovation Academy of Precision Measurement Science and Technology, Chinese Academy of Sciences, Wuhan 430071, China}
\affiliation{Research Center for Quantum Precision Measurement, Guangzhou Institute of Industry Technology, Guangzhou, 511458, China }
\author{Q. Zhong}
\affiliation{Department of Engineering Science and Mechanics, and Materials Research Institute, Pennsylvania State University, University Park, State College, Pennsylvania 16802, USA }
\author{A. Kecebas}
\affiliation{Department of Engineering Science and Mechanics, and Materials Research Institute, Pennsylvania State University, University Park, State College, Pennsylvania 16802, USA }
\author{\c{S}.~K.~\"{O}zdemir}
\email{sko9@psu.edu}
\affiliation{Department of Engineering Science and Mechanics, and Materials Research Institute, Pennsylvania State University, University Park, State College, Pennsylvania 16802, USA }
\author{F.~Zhou}
\email{zhoufei@wipm.ac.cn}
\affiliation{State Key Laboratory of Magnetic Resonance and Atomic and Molecular Physics, Wuhan Institute of Physics and Mathematics, Innovation Academy of Precision Measurement Science and Technology, Chinese Academy of Sciences, Wuhan 430071, China}
\affiliation{Research Center for Quantum Precision Measurement, Guangzhou Institute of Industry Technology, Guangzhou, 511458, China }
\author{H.~Jing}
\email{jinghui73@foxmail.com}
\affiliation{Key Laboratory of Low-Dimensional Quantum Structures and Quantum Control of Ministry of Education, Department of Physics and Synergetic Innovation Center for Quantum Effects and Applications, Hunan Normal University, Changsha 410081, China}
\author{M.~Feng}
\email{mangfeng@wipm.ac.cn}
\affiliation{State Key Laboratory of Magnetic Resonance and Atomic and Molecular Physics, Wuhan Institute of Physics and Mathematics, Innovation Academy of Precision Measurement Science and Technology, Chinese Academy of Sciences, Wuhan 430071, China}
\affiliation{Research Center for Quantum Precision Measurement, Guangzhou Institute of Industry Technology, Guangzhou, 511458, China }
\affiliation{Department of Physics, Zhejiang Normal University, Jinhua 321004, China}

\maketitle

%%%%%%%%%%%%%%%%%%%%%%%%%%%%%%%%%%%%%%%%%%%%Here is the end.
%%% about 150 words %%%

\textbf{
	\\Quantum heat engines and refrigerators are open quantum systems, whose dynamics can be well understood using a non-Hermitian formalism. A prominent feature of non-Hermiticity is the existence of exceptional points (EPs), which has no counterpart in closed quantum systems. It has been shown in classical systems that dynamical encirclement in the vicinity of an EP, whether the loop includes the EP or not, could lead to chiral mode conversion. Here, we show that this is valid also for quantum systems when dynamical encircling is performed in the vicinity of their Liouvillian EPs (LEPs) which include the effects of quantum jumps and associated noise -- an important quantum feature not present in previous works. We demonstrate, using a Paul-trapped ultracold ion, the first chiral quantum heating and refrigeration by dynamically encircling a closed loop in the vicinity of an LEP. We witness the cycling direction to be associated with the chirality and heat release (absorption) of the quantum heat engine (quantum refrigerator). Our experiments have revealed that not only the adiabaticity-breakdown but also the Landau-Zener-St{\"u}ckelberg process play an essential role during dynamic encircling, resulting in chiral thermodynamic cycles. Our observations contributes to further understanding of chiral and topological features in non-Hermitian systems and pave a way to exploring the relation between chirality and quantum thermodynamics.}

\vspace{0.5cm}

%%%%%%%%%%%%%%%%%%%%%%%%%%%%%%%%%%%%%%%%%%%%%%%%%%%%%%%%%%%%%%%%%%%%%%%%%%%%%%%%%%%%%%%%%%%%%%%%%%%%%%%%%%%%%%%%%%%%%%%%%%%%%%%%%%%%%%%%%%%%%%%%%%%%%%%%%%%%%%%%%%%%%%%%%%%%%%%%%%%%%%%%%%%%%%%%%%%%%%%%%%%%%%%%%%%%%%%%%%%%%%%%%%%%%%%%%%%%%%%%%%%%%%%%%%%%
%%% introduction ~ 365 words

\renewcommand\section[1]{
	\textbf{#1}
}
\clearpage
\section{\\Introduction}
\\Quantum heat engines (QHEs), using quantum matter as their working substance, convert the heat energy from thermal reservoirs into useful work. QHEs have been implemented in various microscopic and nanoscopic systems, including single trapped ions and spin ensembles ~\cite{science-352-235,PRL-123-080602,PRL-100-140501,PRL-123-240601,PNAS-111-13786,PRL-125-166802,PRL-122-110601}. Routes to realize QHEs in superconducting circuits~\cite{PRL-97-180402} and quantum optomechanical systems~\cite{PRL-112-150602,PRL-114-183602} have also been proposed. Quantum refrigerators (QRs) are typically achieved by reversing the sequence of the strokes of QHEs~\cite{ARPC-65-365,PRL-108-120602}, removing heat from the cold bath at the expense of external work performed on the system. QRs have been realized with superconducting qubits~\cite{Eur-97-40003,PRB-94-235420}, quantum dots~\cite{PRL-110-256801}, and trapped ions~\cite{Natcommum-10-202,QST-1-015001}.
%\\[1pt]

Another emergent field attracting widespread interest is non-Hermitian dynamics in classical and quantum systems and exotic features associated with non-Hermitian spectral degeneracies known as exceptional points (EPs)~\cite{science-363-42,Natmater-18-783,science-364-878,nature-607-271,science-376-184,NC-13-599,NP-15-1232,prl-126-083604,prl-128-160401}. In contrast to Hermitian spectral degeneracies where eigenvectors associated with degenerate eigenvalues are orthogonal, at an EP both the eigenvalues and the associated eigenvectors become degenerate. The location of the EPs in the parameter space of a system is typically calculated from the system's Hamiltonian which describes the non-unitary coherent evolution of the open system. Such EPs are thus referred to as Hamiltonian EPs (or HEPs).  However, HEPs do not take quantum jumps and the associated noise into account, and therefore does not depict the whole dynamics of an open quantum system. Instead, one should resort to Liouvillian formalism to describe both the non-unitary evolution and the decoherence and hence the quantum jumps. Consequently, the EPs of the system are defined as the eigenvalue degeneracies of Liouvillian superoperators - and thus Liouvillian EPs (or LEPs) \cite{LEP,huber,naka,PRX-Quantum-2-040346,arXiv:2111.04754}.

{\color{black} Open quantum systems exchange energy with external thermal baths, leading to quantum jumps. In the presence of quantum jumps,} the dynamics of QHEs and QRs can be {\color{red}fully described and} well understood using a non-Hermitian framework based on Liouvillian formalism and LEPs{\color{black}, especially for the QHEs based on qubits. Moreover, LEPs introduce unique physical properties to QHEs such as the presence of an LEP can %correspond to the critical decay for QHE to its steady state~\{prxq-2-0403046},
optimise the dynamics of QHEs towards their steady states~\cite{PRX-Quantum-2-040346}, enhance the QHE efficiency~\cite{NC-2022}, and endow topological properties to the QHE~\cite{prl-130-110402}. Nevertheless, LEPs and effects associated with the presence of LEPs have remained largely unexplored in quantum systems and in particular in the field of quantum thermodynamics.}

Dynamically encircling a HEP in a parametric loop has shown to give rise to chiral state transfer due to non-Hermiticity induced non-adiabatic transitions~\cite{PRL-126-170506,nature-537-76,prx-8-021066,prl-124-153903,prl-125-187403,nature-562-86,lsa-8-88,lsa-12-87}. However, recent experimental studies have shown that chiral behaviour can be observed even without encircling an HEP: Any dynamically formed parametric loop in the vicinity of an EP should result in chiral features thanks to the eigenvalue landscape close to the EP~\cite{pra-96-052129,Natcommum-9-4808,pra-102-040201,nature-605-256,nature-537-80}.
{\color{black} Meanwhile, the eigenvalue landscape of an LEP exhibits similar Riemann surfaces, leading to non-trivial state transfer dynamics (e.g., entangled state generation) when the LEP is encircled \cite{arXiv:2111.04754}(\cite{arXiv:2310.11381}). Some experimental studies illustrate the challenge of exploring the counterintuitive chiral behavior in quantum systems without encircling the LEPs~\cite{prl-130-110402,LZS,LZtransition,Sphase}. For example, the topology and landscape of the Riemann surface, along with the trajectory and evolution speed of the dynamical process, significantly influence the results of parametric loops~\cite{prl-130-110402}. These influences result from various aspects, including the phases of the Landau-Zener-St{\"u}ckelberg (LZS) process~\cite{LZS,LZtransition,Sphase}, quantum coherence, net work, and efficiency of the quantum heat engine.}

In this Letter, we experimentally demonstrate {\color{black}chiral behaviour in a qubit system without encircling} LEPs. Namely, we show chiral operation induced by parametric loops in the vicinity of an LEP (without encircling it) in the parameter space of a single trapped ion  configured as a quantum engine for heating and refrigeration. Our work brings together quantum thermodynamics, LEPs, and chiral state transfer due to breakdown of adiabaticity, demonstrating that non-adiabacity and LZS process \cite{LZS,LZtransition,Sphase} are essential to chiral thermodynamic cycles. Our experiment connects, for the first time, the LZS process to chirality in association with LEP-related thermodynamic effects.

\section{\\Results and Discussions}
\textbf{\\Experimental setup.}~ Our experiment is carried out in a single ultracold $^{40}\mathrm{Ca}^{+}$ ion confined in a linear Paul trap (Fig. \ref{Fig1}(a) with more details in~\cite{SA-2-e1600578,njp-19-063032}) whose axial and radial frequencies are, respectively, $\omega_z/2\pi=1.01$ MHz and $\omega_r/2\pi=1.2$ MHz under the pseudo-potential approximation.
{\color{green} }Under an external magnetic field 0.6 mT directed in axial orientation, the ground state $|4^{2}S_{1/2}\rangle$
is split into two hyperfine energy levels while the metastable state $|3^{2}D_{5/2}\rangle$ split into six. As shown in Fig. \ref{Fig1}(b), we label
$|4^{2}S_{1/2}, m_{J}=+1/2\rangle$ as $|g\rangle$, $|3^{2}D_{5/2}, m_{J}=+5/2\rangle$ as $|e\rangle$ and $|4^{2}P_{3/2}, m_{J}=+3/2\rangle$ as $|p\rangle$. After Doppler and resolved sideband cooling of the ion, we reduce the average phonon number of the $z$-axis motional mode of the ion to be much smaller than 1 with the Lamb-Dicke parameter $\sim$ 0.11, which is sufficient to avoid detrimental effects of thermal phonons (e.g., Rabi oscillation offsets). Introducing the dipolar transition $|e\rangle\rightarrow|p\rangle$ by switching on 854-nm laser as explained in \cite{PRR-2-033082, NC-2022},
we reduce this three-level system to an effective two-level system representing a qubit \cite{explain1}. We employ this qubit as the working substance of the QHE and QR.

\begin{figure}[htbp]
\centering
\includegraphics[width=16 cm]{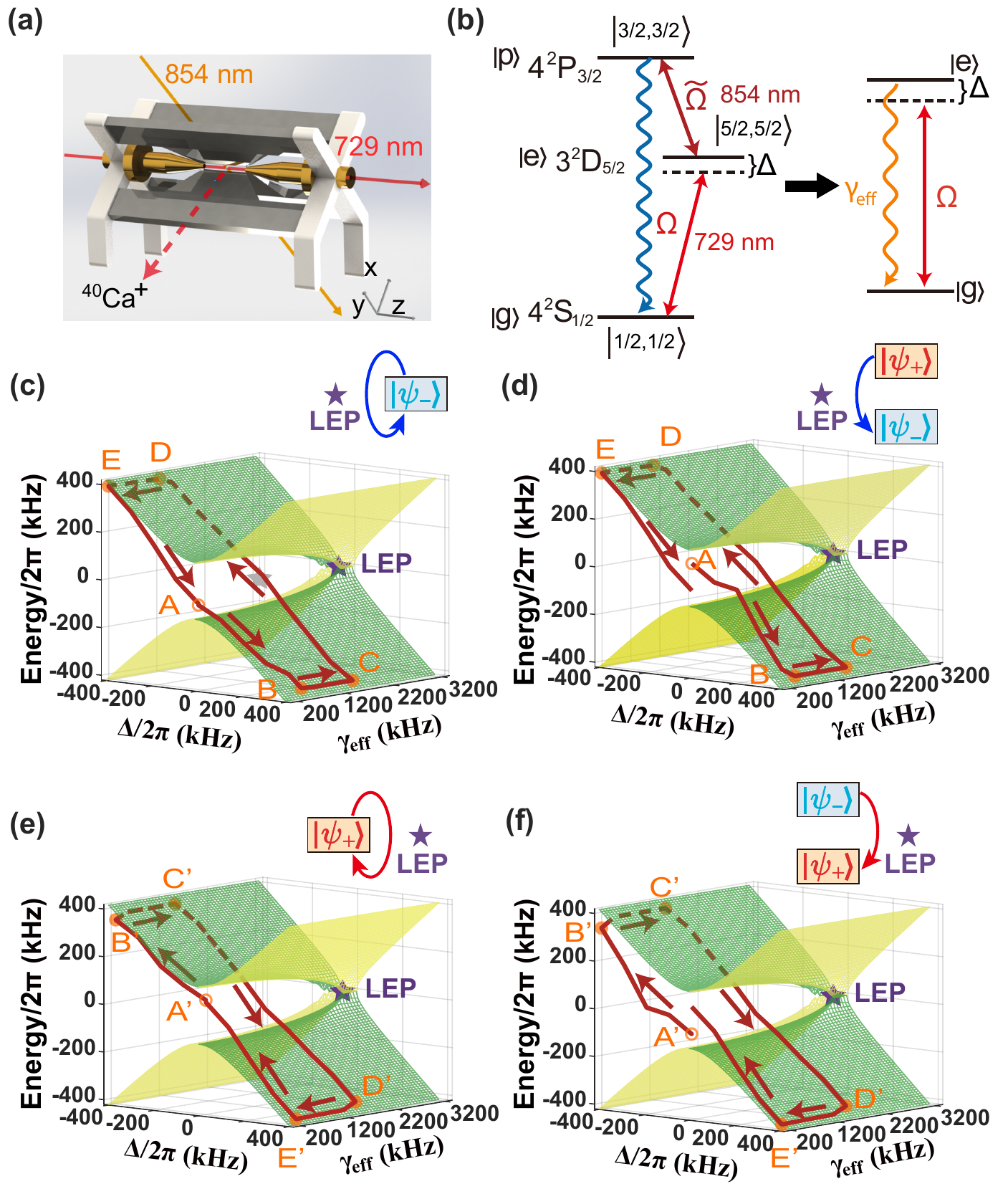}
\caption{Parametric loops in the vicinity of an LEP in the parameter space of a single trapped ion quantum heat engine. (a) The linear Paul trap. (b) Level scheme of the ion, where the solid arrows represent the transitions with Rabi frequencies $\Omega$ and $\tilde{\Omega}$ driven by 729 nm and 854 nm lasers, respectively, and $\Delta$ is the detuning between the energy level and 729 nm laser. The wavy arrow denotes the spontaneous emission with decay rate $\Gamma$. This three-level model can be simplified to an effective two-level system with tunable drive and decay. (c-f) Trajectories without encircling the LEP on the Riemann sheets when completing a clockwise or counterclockwise encirclement in $\Delta\mbox{-}\gamma_{\mathrm{eff}}$ parametric space starting from $|\psi_{+}\rangle$ or $|\psi_{-}\rangle$, where the parameters take the values $\Delta_{\mathrm{min}}/2\pi=-400$ kHz, $\Delta_{\mathrm{max}}/2\pi=400$ kHz, $\gamma_{\mathrm{min}}\approx100$ kHz, $\gamma_{\mathrm{max}}\approx1.45$ MHz. The empty circles represent the starting points. Five orange corner points A, B, C, D, E (A', B', C', D', E') are labeled for convenience of description in the text. The LEPs are labeled by stars.} \label{Fig1}
\end{figure}

The dynamical evolution of this effective two-level model is governed by the Lindblad master equation,
%\begin{equation}
%\begin{array}{cll}
$\dot{\rho}=\mathcal{L}\rho =-i[H_{\mathrm{eff}},\rho]+\frac{\gamma_{\mathrm{eff}}}{2}(2\sigma_{-}\rho\sigma_{+}-\sigma_{+}\sigma_{-}\rho-\rho\sigma_{+}\sigma_{-})$,
%\end{array}
%\end{equation}
where $\mathcal{L}$ is the Liouvillian superoperator, and $\rho$ and $\gamma_{\mathrm{eff}}$ denote the density operator and the effective decay rate
from $|e\rangle$ to $|g\rangle$, respectively. Here the effective Hamiltonian is
%\begin{equation}
%\begin{array}{ccc}
$H_{\mathrm{eff}}=\Delta|e\rangle\langle e|+\frac{\Omega}{2}(|e\rangle\langle g|+|g\rangle\langle e|)$,
%\end{array}
%\end{equation}
where $\Delta$ represents the frequency detuning between the resonance transition and the driving laser while $\Omega$ denotes the Rabi frequency. The eigenvalues of $\mathcal{L}$
at $\Delta=0$ are given by $\lambda_{1}=0, \lambda_{2}=-\gamma_{\mathrm{eff}}/2, \lambda_{3}=(-3\gamma_{\mathrm{eff}}-\xi)/4$ and $\lambda_{4}=(-3\gamma_{\mathrm{eff}}+\xi)/4$, with $\xi=\sqrt{\gamma_{\mathrm{eff}}^2-16\Omega^2}$. It is evident that the eigenvalues $\lambda_{3}$ and $\lambda_{4}$ coalesce when $\gamma_{\mathrm{eff}}=4\Omega$, giving rise to a second order LEP at $\tilde{\lambda}=-3\gamma_{\mathrm{eff}}/4$ \cite{SM}. In the weak coupling situation $\gamma_{\mathrm{eff}}>4\Omega$, both $\lambda_{3}$ and $\lambda_{4}$ are real with a splitting of $\xi/2$. This regime corresponds to the broken phase characterized by a non-oscillatory dynamics accompanied by purely exponential decay \cite{huber,naka}. In the strong coupling regime ($\gamma_{\mathrm{eff}}<4\Omega$), $\lambda_{3}$ and $\lambda_{4}$ are a pair of complex conjugates with the imaginary parts $-\xi/4$ and $\xi/4$, respectively. This regime corresponds to the exact phase characterized by an oscillatory dynamics. Thus LEP acts as a critical damping point of the damped harmonic oscillator which divides the parameter space into a region of oscillatory dynamics (exact phase, $\gamma_{\mathrm{eff}}<4\Omega$) and a region of non-oscillatory dynamics (broken phase, $\gamma_{\mathrm{eff}}>4\Omega$).

{\color{black}
Our QHE and QR cycles differ from their classical counterparts in the definition and implementation of the characteristic thermodynamic quantities. The working substance in our QHE and QR is the qubit defined above. The thermal baths consist of the 729-nm laser irradiation and the actual environment. This working substance carries out work by varying the detuning $\Delta$ in the effective Hamiltonian $H_{\mathrm{eff}}$, in which the increase and decrease of the population %$\rho_{ee}$
in the excited state $|e\rangle$ correspond to heat absorption and heat release
%exothermic and endothermic processes
of the working substance, respectively. As a result, the Rabi interactions in the QHE and QR cycles represent the energy exchange between the working substance and thermal baths. Variations in the effective Hamiltonian lead to LZS process, along with performed work, heat absorption, and heat release ~\cite{prl-130-110402}. }{\color{black}In our experiments, all thermodynamic quantities are acquired from population variations of the qubit and the corresponding tunable parameters \cite{SM}.}
%We maintain a constant Rabi interaction strength $\Omega$ to induce the Riemann surface of our experimental system in Figs. \ref{Fig1}(c)-(f).}

Figures \ref{Fig1}(a,b) present the thermodynamic cycle we carry out, consisting of two iso-decay and two isochoric strokes with a fixed Rabi frequency $\Omega$ as in Ref.~\cite{prl-130-110402}. For the iso-decay strokes, we achieve positive (negative) work by decreasing (increasing) the detuning $\Delta$ while keeping the decay rate $\gamma_{\mathrm{eff}}$ constant. The iso-decay strokes shift the energy difference between the two levels of the working substance and function as the expansion and compression processes for the work done. In the isochoric strokes, our system rapidly reaches the steady state. Then, with the constant values of $\Delta$ and $\Omega$, we increase (or decrease) the population of the excited state for heating (or cooling) by decreasing (increasing) the decay rate $\gamma_{\mathrm{eff}}$. This indicates that the increase (decrease) of the population in the excited state corresponds to the heat absorption (release) from (to) the thermal baths. 

To study the effect of the parametric loops in the vicinity of the LEP on the thermodynamic cycle and the engine performance, we prepare the qubit in the superposition state $|\psi_{+}\rangle=(|e\rangle+|g\rangle)/\sqrt{2}$ or $|\psi_{-}\rangle=(|e\rangle-|g\rangle)/\sqrt{2}$, and then  perform a thermodynamic cycle, corresponding to a loop in the parameter space, by tuning $\gamma_{\mathrm{eff}}$ and $\Delta$ such that the LEP of the system is not encircled. The superposition states $|\psi_{+}\rangle$ and $|\psi_{-}\rangle$ correspond to two eigenstates of $H_{\mathrm{eff}}$ for $\Delta=0$ kHz at the middle point of the expansion and compression {\color{black}processes in the iso-decay} strokes, respectively.Moreover, to prevent our operations from affecting the vibrational modes of the ion, leading to unexpected heating noises, we restrict $|\Delta|$ to be smaller than half of the vibrational frequency.

Starting from the mid-points of these strokes requires five strokes to execute a clockwise (CW) or a counterclockwise (CCW) loop. After the loop (CW or CCW) is completed, we compare the final state $\rho$ of the {\color{black}qubit} with the initial state $|\psi_{+}\rangle$ or $|\psi_{-}\rangle$ by calculating the fidelity $\langle\psi_{+}|\rho|\psi_{+}\rangle$ or $\langle\psi_{-}|\rho|\psi_{-}\rangle$. Figures \ref{Fig1}(c) and \ref{Fig1}(d) show the evolution of the system for two counterclockwise loops that start at the initial states $|\psi_{-}\rangle$ and $|\psi_{+}\rangle$, respectively, and are completed without encircling the LEP. We first implement an iso-decay compression stroke from A to B by increasing the detuning $\Delta$ linearly from 0 to its maximum value $\Delta_{\mathrm{max}}$ while keeping $\gamma_{\mathrm{eff}}$ at its minimum value of $\gamma_{\mathrm{min}}$. The second stroke is an isochoric cooling stroke from B $(\Delta_{\mathrm{max}},\gamma_{\mathrm{min}})$ to C $(\Delta_{\mathrm{max}}, \gamma_{\mathrm{max}})$, which is realized by increasing $\gamma_{\mathrm{eff}}$ from $\gamma_{\mathrm{min}}$ to $\gamma_{\mathrm{max}}$ with the detuning remaining constant at $\Delta_{\mathrm{max}}$. Then we execute the third stroke, i.e., an iso-decay expansion from C $(\Delta_{\mathrm{max}},\gamma_{\mathrm{max}})$ to D $(\Delta_{\mathrm{min}},\gamma_{\mathrm{max}})$, by decreasing $\Delta$ from $\Delta_{\mathrm{max}}$ to $\Delta_{\mathrm{min}}$ while keeping $\gamma_{\mathrm{eff}}=\gamma_{\mathrm{max}}$. The fourth stroke is an isochoric heating from D $(\Delta_{\mathrm{min}},\gamma_{\mathrm{max}})$ to E $(\Delta_{\mathrm{min}},\gamma_{\mathrm{min}})$ executed by decreasing $\gamma_{\mathrm{eff}}$ from $\gamma_{\mathrm{max}}$ to $\gamma_{\mathrm{min}}$ with $\Delta$ fixed at $\Delta_{\mathrm{min}}$. The final step is an iso-decay compression stroke from E $(\Delta_{\mathrm{min}},\gamma_{\mathrm{min}})$ back to A$(0,\gamma_{\mathrm{min}})$ executed by increasing $\Delta$ from $\Delta_{\mathrm{min}}$ to 0 while $\gamma_{\mathrm{eff}}$ is kept fixed at $\gamma_{\mathrm{min}}$. The clockwise loops depicted in Figs. \ref{Fig1}(e) and \ref{Fig1}(f) are also executed in five strokes with
the initial states prepared in $|\psi_{+}\rangle$ and $|\psi_{-}\rangle$ but in reverse order of the process described above for counterclockwise loops.  It is clearly seen that regardless of the initial state that the loops start from, the system always ends up at the final state $|\psi_+\rangle$ for a clockwise loop and at the final state $|\psi_-\rangle$ for a counterclockwise loop. This observation that the execution direction of the loop determines the final state is a signature of the chiral behavior in our system. This process is often referred to as asymmetric mode conversion or chiral state transfer \cite{prx-8-021066,nature-537-80,LZS}.

\begin{figure*}[htbp]
\centering
\includegraphics[width=17 cm]{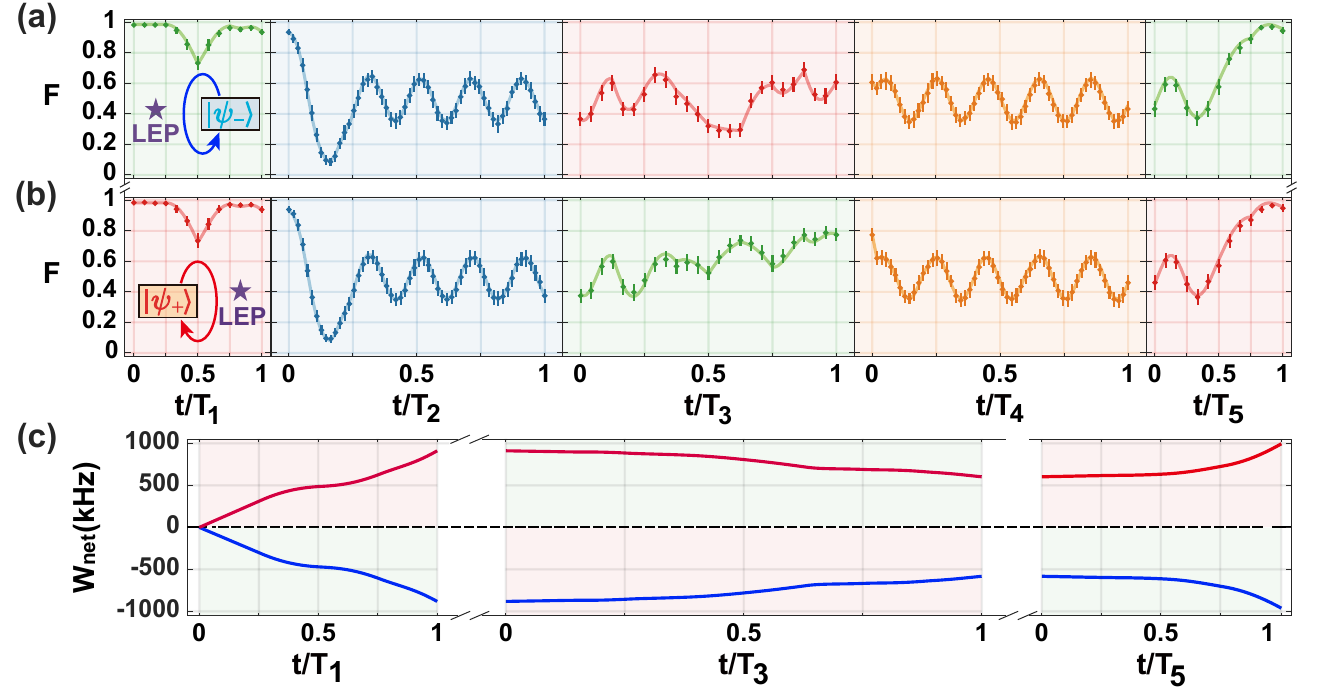}
\caption{Closed loops that do not exhibit mode conversion. (a) Counterclockwise loop starting from $|\psi_{-}\rangle$. (b) Clockwise loop starting from $|\psi_{+}\rangle$. Evolution of the system's state along the trajectory of the closed loop is characterized by the fidelity $\langle\psi_{-}|\rho(t)|\psi_{-}\rangle$ in (a) and the fidelity $\langle\psi_{+}|\rho(t)|\psi_{+}\rangle$ in (b). The circles and error bars denote, respectively, the average and standard deviations of 10000 measurements. The solid curves connecting the dots are obtained by simulating master equations. (c) Blue and red solid curves represent, respectively, the time evolution of the net work in the counterclockwise and clockwise loops, where only three of the strokes associated with doing work are depicted.
%Regions with different colors represent, respectively, different strokes of the heat engine cycle and refrigerator cycles, with the orange and blue corresponding to isochoric heating and isochoric cooling strokes respectively and the red and green denoting iso-decay expansion and iso-decay compression strokes, respectively.
Durations of the five strokes are $T_{1}=T_{5}=6$ $\mu$s, $T_{3}=12$ $\mu$s, and $T_{2}=T_{4}=150$ $\mu$s. Other parameters are $\Omega/2\pi=$ 120 kHz, $\Delta_{\mathrm{min}}/2\pi=-400$ kHz, $\Delta_{\mathrm{max}}/2\pi=400$ kHz, $\gamma_{\mathrm{min}}\approx0$ kHz, and $\gamma_{\mathrm{max}}\approx1.45$ MHz.} \label{Fig2}
\end{figure*}

\textbf{Quantum heating and cooling.} Figure 2 displays experimental results of a counterclockwise loop starting at $|\psi_{-}\rangle$ and a clockwise loop starting at $|\psi_{+}\rangle$, corresponding to the schemes in Fig. \ref{Fig1}(c,e). These loops do not encircle the LEP of the system. In Fig. 2(b), we prepare the {\color{black}qubit} in the initial state $|\psi_{+}\rangle$ with fidelity $\langle\psi_{+}|\rho_{A}|\psi_{+}\rangle=0.985$ which is lower than 1 due to imperfections during the state preparation. The system returns to $|\psi_{+}\rangle$ after completion of the loop. Figure 2(a) exhibits the results of the counterclockwise loop starting at $|\psi_{-}\rangle$. The state of the system evolves back to the initial state $|\psi_{-}\rangle$ when the loop is completed. The two situations correspond, respectively, to the QR and QHE cycles, as elucidated later. For the counterclockwise encirclement in Fig. 2(a), the system is initialized to $|\psi_{-}\rangle=(|e\rangle-|g\rangle)/\sqrt{2}$. The Landau-Zener transition occurs in the first stroke with the detuning $\Delta$ varied from 0 to $\Delta_{\mathrm{max}}/2\pi=400$ kHz, and then the system evolves to a nearly steady state in the second stroke due to the increase of the decay rate with a large detuning $\Delta/2\pi=400$ kHz. In the third stroke, the LZS process occurs, resulting in the population oscillation. After the fourth stroke the system evolves to another nearly steady state as a result of the large detuning $\Delta/2\pi=-400$ kHz and slow-varying dissipation. Finally, a Landau-Zener transition occurs again in the fifth stroke with the detuning $\Delta$ tuned from $\Delta_{\mathrm{min}}/2\pi=-400$ kHz to 0, resulting in the final state $|\psi_{-}\rangle=(|e\rangle-|g\rangle)/\sqrt{2}$~\cite{LZtransition,Sphase}. In contrast, for the clockwise loop from $|\psi_{+}\rangle=(|e\rangle+|g\rangle)/\sqrt{2}$, the system evolves to a steady state after the fourth stroke and experiences a LZS with the detuning $\Delta$ tuned from $\Delta_{\mathrm{max}}/2\pi=400$ kHz to 0, accumulating a St{\"u}ckelberg phase~\cite{Sphase} and thus leading to the final state $|\psi_{+}\rangle=(|e\rangle+|g\rangle)/\sqrt{2}$.

Since the system evolves along a closed trajectory, we can evaluate the net work using $W_{1}=\int_{0}^{t}\rho(t)dH(t)$, where $\rho(t)$ describes the state of the two-level system governed by $H(t)=\Delta(t)|e\rangle\langle e|/2$. For the clockwise encirclement in Fig. 2(b), since the isochoric strokes produce no work, here we only consider the first, third, and fifth strokes with the first and fifth strokes executing expansion and the third stroke executing compression. Moreover, the bath performs work on the system in the iso-decay compression stroke (third stroke) and the system produces work during the two iso-decay expansion strokes. The mean population in $|e\rangle$ in the two expansion strokes is higher than that in the compression stroke, implying a positive net work, and thus the system behaves as the QHE. On the contrary, the counterclockwise loop in Fig. \ref{Fig2}(a) leads to negative net work due to the higher mean population in $|e\rangle$ in the two compression strokes than that in the expansion stroke. This makes the system perform as a QR. Thus, we conclude that the loops that do not result in asymmetric mode conversion (CW loop starting at $|\psi_{+}\rangle$ and CCW loop starting at $|\psi_{-}\rangle$) performs as a QHE (Fig. 2(b)) or a QR (Fig. 2(a)).

\begin{figure*}[htbp]
\centering
\includegraphics[width=17cm]{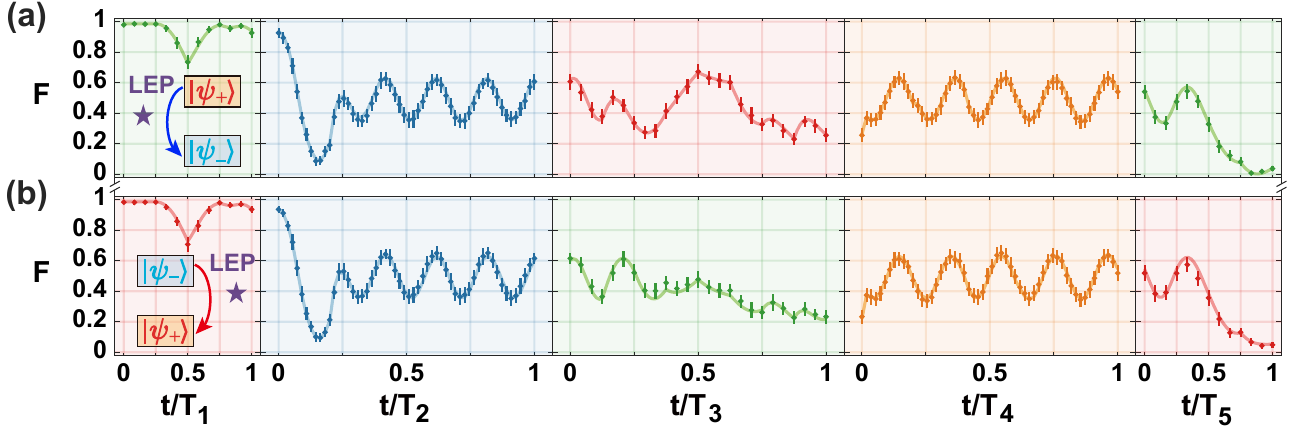}
\caption{Closed loops that lead to mode conversion. (a) Counterclockwise loop starting at $|\psi_{+}\rangle$ and ending at $|\psi_{-}\rangle$. (b) Clockwise loop starting at $|\psi_{-}\rangle$ and ending at $|\psi_{+}\rangle$. Evolution of the system's state along the closed loop trajectory is characterized by the fidelity $\langle\psi_{+}|\rho(t)|\psi_{+}\rangle$ in (a) and the fidelity $\langle\psi_{-}|\rho(t)|\psi_{-}\rangle$ in (b). The circles and error bars respectively denote the average and standard deviations of 10000 measurements. The solid curves are obtained by simulating master equations. Durations of the five strokes are $T_{1}=T_{5}=6$ $\mu$s, $T_{3}=12$ $\mu$s, and $T_{2}=T_{4}=150$ $\mu$s. Other parameters are $\Omega/2\pi=$ 120 kHz, $\Delta_{\mathrm{min}}/2\pi=-400$ kHz, $\Delta_{\mathrm{max}}/2\pi=400$ kHz, $\gamma_{\mathrm{min}}\approx0$ kHz, and $\gamma_{\mathrm{max}}\approx1.45$ MHz.} \label{Fig3}
\end{figure*}

{\color{black} We mention that Rabi frequency $\Omega$ remains constant in our implementation of QHE and QR cycles, indicating that the working substance keeps interacting with the 729-nm laser. The heating and cooling processes are accomplished by modulating the detuning and/or the decay rate. For example, the working substance reaches the thermal equilibrium with the thermal baths by changing $\gamma_{\mathrm{eff}}$ while keeping a constant value of $\Delta$~\cite{pre-76-031105,pr-697-1,PRL-125-166802}. }

\textbf{Chiral dynamics.} The chiral behavior and asymmetric mode conversion appear for counterclockwise loop starting at $|\psi_{+}\rangle$ and the clockwise loop starting at $|\psi_{-}\rangle$, as witnessed in Fig. 3. In both of these two cases, the system cannot return to the starting state after the loop is completed. Together with the observation depicted in Fig. 2(a), we conclude that the final states (i.e., the end points of the loops) depend only on the encircling direction, not on the initial state (i.e., the starting point of the loops): A CW (or CCW) loop in the vicinity of the LEP but not encircling it ends up at the final state $|\psi_{+}\rangle$ (or $|\psi_{-}\rangle$) regardless of whether the initial state is $|\psi_{+}\rangle$ or $|\psi_{-}\rangle$.

%We notice that the Landau-Zener transition in the fifth stroke
It is evident that, whether the system evolves back to the starting point or not is essentially associated with the LZS process and the breakdown of the adiabaticity when executing the closed loop trajectory. This is reflected in the fifth stroke of the loop when a state transfer occurs after experiencing a Landau-Zener transition (Figs. 2 and 3).For the clockwise encirclement starting from $|\psi_{+}\rangle$ or the counterclockwise encirclement from $|\psi_{-}\rangle$, the Landau-Zener transition in the fifth stroke determines if the closed thermodynamic cycle works as the chiral QHE or QR. In contrast, for the clockwise loop starting at $|\psi_{-}\rangle$ or the counterclockwise loop at $|\psi_{+}\rangle$, this Landau-Zener transition determines the chiral behavior.  In other words, when the system starts at the state $|\psi_{+}\rangle$, a clockwise loop in the vicinity of (not encircling) the LEP ends in the same state whereas a counterclockwise loop ends at the orthogonal state $|\psi_{-}\rangle$. So the system acts as a QHE for the clockwise loop and as a mode converter for the counterclockwise loop. Similarly, when the system starts at the state $|\psi_{-}\rangle$, a clockwise loop in the vicinity of (not encircling) the LEP ends in the orthogonal state $|\psi_{+}\rangle$ whereas a counterclockwise loop ends at the same state $|\psi_{-}\rangle$. So the system acts as a QR for the counterclockwise loop and as a mode converter for the clockwise loop.

{\color{black} Here we emphasize that, the LZ process in the fifth stroke is the most essential to the chiral behavior and asymmetric mode conversion
. When the evolution time of the fifth stroke is too short, our system works in the non-adiabatic regime and lacks sufficient time to follow the variation of the detuning, resulting in non-adiabatic transitions between two eigenstates via phase accumulation through the LZS process~\cite{LZS,LZtransition,Sphase}. These transitions would prevent our system from exhibiting chiral behavior and asymmetric mode conversion. In contrast, when the evolution time of the fifth stroke is too long, our system works in the adiabatic regime and smoothly follows the variation of the detuning. Nevertheless, the evolution duration is restricted by the decoherence time, as clarified with the numerical simulation results shown in Fig. S.8.}

Although the displayed loops exclude the LEP, our numerical simulations suggest a strong relevance of the observed chiral behavior to the  existence of the LEP. If the loop is too far away from the LEP, i.e., $\gamma_{\rm{eff}}$ is too small, no chirality would appear \cite{SM}. However, if the system parameters are chosen such that the loops are closer to the LEP, but not encircling it, the populations of $|\psi_{+}\rangle$ and $|\psi_{-}\rangle$ interchange for the clockwise loop starting at $|\psi_{-}\rangle$ and the counterclockwise loop starting at $|\psi_{+}\rangle$ \cite{SM}. Moreover, since the loops do not encircle the LEP, no topological phase transition is involved in system's response. We note that an essential ingredient for the observed chirality is the transition of the system between two Riemann sheets when the parameters are tuned to form a closed loop. Considering the system state evolves only within a single Riemann sheet, we see the encircling direction determines whether the chirality exists or not. For example, Figs. S5 and S6 in \cite{SM} show that chiral behaviour appears only in the counterclockwise and clockwise encirclements, respectively. Otherwise the final state neither returns to the initial state nor has state convention between $|\psi_{+}\rangle$ and $|\psi_{-}\rangle$. This phenomena can be called non-reciprocal chirality, i.e., unidirectional chirality.

As our chiral behavior and asymmetric mode conversion result from the LEP rather than the Hamiltionian EP~\cite{lsa-8-88,lsa-12-87,PRL-126-170506,nature-537-76,prx-8-021066,prl-124-153903,prl-125-187403,nature-562-86}, we have fully captured the quantum dynamics including the quantum jumps and associated noises, significantly extending the realization conditions of chiral behavior and asymmetric mode conversion, and considerably reducing the experimental difficulties in quantum control and measurement. {\color{black} Realizing chiral behavior without encircling an LEP helps reduce the parameter space needed to steer the system. For example, encircling the LEP required varying $\gamma_{\mathrm{max}}$ to values larger than 4$\Omega$ [33]; however, observing chiral behavior without encircling the LEP requires varying $\gamma_{\mathrm{max}}$ to values less than 2$\Omega$. This helps keeping our system in the quantum regime for achieving the chiral behavior and asymmetric mode conversion, and also opens up possibilities for a more focused and extended exploration of the physical properties associated with the LEP.}
Moreover, the combination of the adiabaticity breakdown~\cite{lsa-8-88} and the Landau-Zener-St{\"u}ckelberg phase~\cite{LZtransition,Sphase} leads to the presented chirality, whose physical mechanics is distinct from the one resulting from spontaneous chiral symmetry breaking~\cite{prl-118-033901}.  Furthermore, our observed chiral behavior and asymmetric mode conversion strongly depend on the LEPs resulting from quantum jumps. Characterized by quantum jumps, the above phenomena in our experimental system are inherently quantum. In contrast, without quantum jumps, our experimental system would return to Hamiltonian EPs~\cite{prl-126-083604,pra-103-l020201}, whose chiral behavior and asymmetric mode conversion have been theoretically predicted~\cite{pra-96-052129} and experimentally observed with waveguides~\cite{pra-102-040201} and fibers~\cite{nature-605-256}.

\section{\\Conclusion}
\\In conclusion, we have experimentally demonstrated, for the first time, a chiral behavior in a single trapped-ion system without dynamically encircling its LEP. We show clearly that asymmetric mode conversion is directly related to the topological landscape of the Riemann surfaces and not necessarily to encircling an LEP of this quantum system, supporting previous reports for classical systems. Our experiments may open up new avenues in understanding the chiral and topological behaviors in non-Hermitian systems and bridging chirality and quantum thermodynamics.

\section{Data availability}\\
The data illustrated in the figures within this paper are available from the corresponding authors upon request. % Source data are provided with this paper.

\section{Code availability}\\
The custom codes used to generate the results presented in this paper are available from the corresponding authors upon request.

\section{\\Acknowledgments}\\This work was supported by Special Project for Research and Development in Key Areas of
Guangdong Province under Grant No. 2020B0303300001, by National Natural Science Foundation of China under Grant Nos. U21A20434, 92265107, 12074390, 11835011, by Key Lab of Guangzhou for Quantum Precision Measurement under Grant No.202201000010, by Postdoctoral Science Foundation of China under Grant No. 2022MT10881, and  by Nansha Senior Leading Talent Team Technology Project under Grant No. 2021CXTD02. \c{S}.K.O acknowledges the support from Air Force Office of Scientific Research (AFOSR) Multidisciplinary University Research Initiative (MURI) Award No. FA9550-21-1-0202 and the AFOSR Award No. FA9550-18-1-0235.  \\

\section{\\Author contributions}\\
HJ and MF conceived the idea and designed the experiments; JTB and FZ performed the experiments with help from JCL,BW, WQD, WFY and JWZ; GYD and JQZ analyzed the data with help form QZ, AK, and LC . Theoretical background and simulations were provided by GYD and JQZ. All authors contributed to the discussions and the interpretations of the experimental and theoretical results. MF, HJ, and SKO wrote the manuscript with inputs from all authors.  \\

\section{\\Competing interests}\\
\\The authors declare that they have no competing interests.

%\clearpage
%\section{\\References:}
%\bibliography{myref}

\begin{thebibliography}{99}
\bibitem{science-352-235}  Ro$\beta$nagel J. , Dawkins S. T. , Tolazzi  K. N. , Abah O. , Lutz E. , Schmidt-Kaler F. , \& Singer K. ,
A single-atom heat engine,
{\it Science\/} \textbf{352}, 325 (2016).

\bibitem{PRL-123-080602} Lindenfels D. von. , Gr\"{a}b O. , Schmiegelow C. T. , Kaushal V. , Schulz J. , Mitchison M. T. , Goold J. , Schmidt-Kaler F. ,\& Poschinger U. G. ,
Spin Heat Engine Coupled to a Harmonic-Oscillator Flywheel,
{\it Phys. Rev. Lett.\/} {\bf 123}, 080602 (2019).

\bibitem{PRL-100-140501} Ryan C. A. , Moussa O. , Baugh J. , \& Laflamme R. ,
Spin Based Heat Engine: Demonstration of Multiple Rounds of Algorithmic Cooling,
{\it Phys. Rev. Lett.\/} {\bf 100}, 140501 (2008).

\bibitem{PRL-123-240601} Peterson J. P. S. , Batalh$\tilde{a}$o T. B. , Herrera M. , Souza A. M. , Sarthour R. S. , Oliveira I. S. , \& Serra R. M. , Experimental Characterization of a Spin Quantum Heat Engine,
{\it Phys. Rev. Lett.\/} {\bf 123}, 240601 (2019).

\bibitem{PNAS-111-13786} Koski J. V. , Maisi V. F. , Pekola J. P. , \& Averin D. V. ,
Experimental realization of a Szilard engine with a single electron,
{\it Proc. Natl. Acad. Sci. U.S.A.\/} {\bf 111}, 13786 (2014).

\bibitem{PRL-125-166802} Ono K. , Shevchenko S. N. , Mori T. , Moriyama S. , \& Nori F. ,
Analog of a Quantum Heat Engine Using a Single-Spin Qubit,
{\it Phys. Rev. Lett.\/} {\bf 125}, 166802 (2020).

\bibitem{PRL-122-110601} Klatzow J. , Becker J. N. , Ledingham P. M. , Weinzetl C. , Kaczmarek K. T. ,
Saunders D. J. , Nunn J. , Walmsley I. A. , Uzdin R. , \& Poem E. ,
Experimental Demonstration of Quantum Effects in the Operation of Microscopic Heat Engines,
{\it Phys. Rev. Lett.\/} {\bf 122}, 110601 (2019).

\bibitem{PRL-97-180402} Quan H. T. , Wang Y. D. , Liu Y. X. , Sun C. P. , \& Nori F. ,
Maxwell Demon Assisted Thermodynamic Cycle in Superconducting Quantum Circuits,
{\it Phys. Rev. Lett.\/} {\bf 97}, 180402 (2006).

\bibitem{PRL-112-150602} Zhang K. , Bariani F. , \& Meystre P. ,
Quantum optomechanical heat engine,
{\it Phys. Rev. Lett.\/} {\bf 112}, 150602 (2014).

\bibitem{PRL-114-183602} Dechant A. , Kiesel N. , \& Lutz E. ,
All-Optical Nanomechanical Heat Engine,
{\it Phys. Rev. Lett.\/} {\bf 114}, 183602 (2015).

\bibitem{ARPC-65-365} Kosloff R. , \& Levy A. ,
Quantum Heat Engines and Refrigerators: Continuous Devices,
{\it Annu. Rev. Phys. Chem.\/} {\bf  65}, 365 (2014).

\bibitem{PRL-108-120602} Mari A. , \& Eisert J. ,
Cooling by Heating: Very Hot Thermal Light Can Significantly Cool Quantum Systems,
{\it Phys. Rev. Lett.\/} {\bf 108}, 120602 (2012).

\bibitem{Eur-97-40003} Levy A. , \& Kosloff R. ,
Quantum Absorption Refrigerator,
{\it Phys. Rev. Lett.\/} {\bf 108}, 070604 (2012).

\bibitem{PRB-94-235420} Tan K. Y. , Partanen M. , Lake R. E. , Govenius J. , Masuda S. , \& M\"{o}tt\"{o}nen M. ,
Quantum-circuit refrigerator.
{\it Nat. Commun.\/} {\bf 8}, 15189 (2017).

\bibitem{PRL-110-256801} Venturelli D. , Fazio R. , \& Giovannetti V. ,
Minimal Self-Contained Quantum Refrigeration Machine Based on Four Quantum Dots,
{\it Phys. Rev. Lett.\/} {\bf 110}, 256801 (2013).

\bibitem{Natcommum-10-202} Maslennikov G. , Ding S. Q. , Hablutzel R. , Gan J. ,
Roulet A. , Nimmrichter S. , Dai J. , Scarani V. , \& Matsukevich D. ,
Quantum absorption refrigerator with trapped ions,
{\it Nat. Commun.\/} {\bf 10}, 202 (2019).

\bibitem{QST-1-015001} Mitchison M. T. , Huber M. , Prior J. , Woods M. P. , \& Plenio M. B. ,
Realising a quantum absorption refrigerator with an atom-cavity system,
{\it Quantum Sci. Technol.\/} {\bf 1}, 015001 (2016).

\bibitem{science-363-42} Miri M.-A. , \& Al\`{u} A. ,
Exceptional points in optics and photonics,
{\it Science\/} {\bf 363}, 42 (2019).

\bibitem{Natmater-18-783} \"{O}zdemir \c{S}. K. , Rotter S. , Nori F. , \& Yang L. ,
Parity-time symmetry and exceptional points in photonics,
{\it Nat. Mater.\/} \textbf{18}, 783 (2019).

\bibitem{science-364-878} Wu Y. , Liu W. , Geng J. , Song X. , Ye X. , Duan C.-K. , Rong X. , \& Du J. ,
Observation of parity-time symmetry breaking in a single-spin system,
{\it Science\/} \textbf{364}, 878 (2019).


\bibitem{nature-607-271} Patil Y. S. S. , H{\"o}ller J. , Henry P. A. , Guria C. ,
Zhang Y. , Jiang L. , Kralj N. , Read N. , \& Haris J. G. E. ,
Measuring the knot of non-Hermitian degeneracies and non-commuting braids,
{\it Nature (London)\/} {\bf 607}, 271 (2022).

\bibitem{science-376-184} Ergoktas M. S. , Soleymani S. , Kakenov N. , Wang K. Y. , Smith T. B. , Bakan G. , Balci S. ,
Principi A. , Novoselov K. S. , {\"O}zdemir \c{S}. K. , Kocabas C. ,
Topological engineering of terahertz light usingelectrically tunable exceptional point singularities,
{\it Science\/} \textbf{376}, 184(2022).

\bibitem{NC-13-599} Soleymani S. , Zhong Q. , Mokim M. , Rotter S. , Ganainy R. E. , \& \"{O}zdemir \c{S}. K. ,
Chiral and degenerate perfect absorption on exceptional surfaces,
{\it Nat. Commun.} {\bf 13}, 599 (2022).



\bibitem{NP-15-1232} Naghiloo M. , Abbasi M. , Joglekar Y. N. , \&  Murch K. W. ,
Quantum state tomography across the exceptional point in a single dissipative qubit,
{\it Nat. Phys.} {\bf 15}, 1232 (2019).

\bibitem{prl-126-083604} Ding L. Y. , Shi K. Y. , Zhang Q. X. , Shen D. N. , Zhang X. , \& Zhang W. ,
Experimental Determination of $\mathcal{PT}$-Symmetric Exceptional Points in a Single Trapped Ion,
{\it Phys. Rev. Lett.\/} {\bf 126}, 083604 (2021).

\bibitem{prl-128-160401} Abbasi M. , Chen W. J. , Naghiloo M. , Joglekar Y. N. ,\& Murch K. W. ,
Topological Quantum State Control through Exceptional-Point Proximity,
{\it Phys. Rev. Lett.\/} {\bf 128}, 160401 (2022).

\bibitem{LEP} Minganti F. , Miranowicz A. , Chhajlany R. W. , \& Nori F. ,
Quantum exceptional points of non-Hermitian Hamiltonians and Liouvillians: The effects of quantum jumps,
{\it Phys. Rev. A\/} {\bf 100}, 062131 (2019).

\bibitem{huber} Huber J. , Kirton P. , Rotter S. , \& Rabl P. ,
Emergence of $\mathcal{PT}$-symmetry breaking in open quantum systems,
{\it Sci-Post Phys.\/} {\bf 9}, 52 (2020).

\bibitem{naka} Nakanishi Y. , \& Sasamoto T. ,
PT phase transition in open quantum systems with Lindblad dynamics,
{\it Phys. Rev. A} {\bf 105}, 022219 (2020).

\bibitem{PRX-Quantum-2-040346} Khandelwal S. , Brunner N. , \& Haack G. ,
Signatures of Liouvillian Exceptional Points in a Quantum Thermal Machine,
{\it Phys. Rev. X. Quantum.\/} {\bf 2}, 040346 (2021).

\bibitem{arXiv:2111.04754} Chen W. J. , Abbasi M. , Ha B. , Erdamar S. , Joglekar Y. N. , \& Murch K. W. ,
Decoherence Induced Exceptional Points in a Dissipative Superconducting Qubit,
{\it Phys. Rev. Lett.\/} {\bf 128}, 110402 (2022).

\bibitem{NC-2022} Zhang J.-W. , Zhang J.-Q. , Ding G.-Y. , Li J.-C. , Bu J.-T. , Wang B. ,
Yan L.-L. , Su S.-L. , Chen L. , Nori F. , \"{O}zdemir \c{S}. K. , Zhou F. , Jing H. , \& Feng M. ,
Dynamical Control of Quantum Heat Engines Using Exceptional Points,
{\it Nat. Commun.} {\bf 13}, 6225 (2022).

\bibitem{prl-130-110402} Bu J.-T. , Zhang J.-Q. , Ding G.-Y. , Li J.-C. , Zhang J.-W. , Wang B. ,
Ding W.-Q., Yuan W.-F , Chen L. , Nori F. , \"{O}zdemir \c{S}. K. , Zhou F. , Jing H. , \& Feng M. ,
Dynamical Control of Quantum Heat Engines Using Exceptional Points,
{\it Phys. Rev. Lett.} {\bf 130}, 110402 (2023).


\bibitem{PRL-126-170506} Liu W. , Wu Y. , Duan C.-K. , Rong X. ,\& Du J. ,
Dynamically Encircling an Exceptional Point in a Real Quantum System,
{\it Phys. Rev. Lett.\/} {\bf 126}, 170506 (2021).

\bibitem{nature-537-76} Doppler J. , Mailybaev A. A. , Rabl P. , Moiseyev N. , \& Rotter S. ,
Dynamically encircling an exceptional point for asymmetric mode conversion,
{\it Nature (London)\/} {\bf 537}, 76 (2016).

\bibitem{prx-8-021066} Zhang X.-L. , Wang S. , Hou B. , \& Chan C. T. ,
Dynamically Encircling Exceptional Points: In situ Control of Encircling Loops and the Role of the Starting Point,
{\it Phys. Rev. X \/} {\bf 8}, 021066 (2018).

\bibitem{nature-562-86} Yoon J. W. , Choi Y. , Hahn C. , Kim G. , Song S. H. , Yang K.-Y. ,
Lee J. Y. , Kim Y. , Lee C. S. , Shin J. K. , Lee H.-S. , \&  Berini P. ,
Time-asymmetric loop around an exceptional point over the full optical communications band,
{\it Nature (London)\/} {\bf 562}, 86 (2018).

\bibitem{lsa-8-88} Zhang X. L. , Jiang T. , and Chan C. T. ,
Dynamically encircling an exceptional point in anti-parity-time symmetric systems: asymmetric mode switching for symmetry-broken modes,
{\it Light Sci. Appl.} {\bf 8}, 88 (2019).

\bibitem{prl-124-153903} Liu Q. , Li S. , Wang B. , Ke S. , Qin C. , Wang K. , Liu W. , Gao D. , Berini P. , \& Lu P. ,
Efficient Mode Transfer on a Compact Silicon Chip by Encircling Moving Exceptional Points,
{\it Phys. Rev. Lett.\/} {\bf 124}, 153903 (2020).

\bibitem{prl-125-187403} Li A. , Dong J. , Wang J. , Cheng Z. , Ho J. S. ,
Zhang D. , Wen J. , Zhang X.-L. , Chan C. T. , Al\'{u} A. , Qiu C.-W. , \& Chen L. ,
Hamiltonian Hopping for Efficient Chiral Mode Switching in Encircling Exceptional Points,
{\it Phys. Rev. Lett.\/} {\bf 125}, 187403 (2020).

\bibitem{lsa-12-87} Baek S. , Park S. H. , Oh D. , Lee K. , Lee S. , Lim H. , Ha T. , Park H. S. , Zhang S. , Yang L. , Min B. , and Kim T. T. ,
Non-Hermitian chiral degeneracy of gated graphene metasurfaces,
{\it Light Sci. Appl.} {\bf 12}, 87 (2023).

\bibitem{pra-96-052129} Hassan A. U. , Galmiche G. L. , Harari G. , LiKamWa P. , Khajavikhan M. , Segev M. ,\& Christodoulides D. N. ,
Chiral state conversion without encircling an exceptional point,
{\it Phys. Rev. A\/} {\bf 96}, 052129 (2017).

\bibitem{Natcommum-9-4808} Zhong Q. , Khajavikhan M. , Christodoulides D. N. , \& El-Ganainy R. ,
Winding around non-Hermitian singularities,
{\it Nat. Commun.\/} {\bf 9}, 4808 (2018).

\bibitem{pra-102-040201} Feilhauer J. , Doppler J. , Mailybaev A. A. , B{\"o}hm J. , Kuhl U. , Moiseyev N. ,\& Rotter S. ,
Encircling exceptional points as a non-Hermitian extension of rapid adiabatic passage,
{\it Phys. Rev. A\/} {\bf 102}, 040201 (2020).

\bibitem{nature-605-256} Nasari H. , Lopez-Galmiche G. , Lopez-Aviles H. E. , Schumer A. ,
Hassan A. U. , Zhong Q. , Rotter S. , LikamWa P. , Christodoulides D. N. , \& Khajavikhan M. ,
Observation of chiral state transfer without encircling an exceptional point,
{\it Nature \/} {\bf 605}, 256 (2022).

\bibitem{nature-537-80} Xu H. , Mason D. , Jiang L. Y. , \& Harris J. G. E. ,
Topological energy transfer in an optomechanical system with exceptional points,
{\it Nature (London)\/} {\bf 537}, 80 (2016).

\bibitem{arXiv:2310.11381} Khandelwal S., Chen W., Murch K. W., \& Haack G.
Chiral Bell-state transfer via dissipative Liouvillian dynamics,
{\it arXiv:2310.11381\/} (2023).


\bibitem{LZS} Shevchenko S. N. , Ashhab S. , \& Nori F. ,
Landau-Zener-St{\"u}ckelberg interferometry,
{\it Phys. Rep.} {\bf 492}, 1 (2010).

\bibitem{LZtransition} Landau L. ,
On the theory of transfer of energy at collisions II,
{\it Phys. Z. Sowjetunion} {\bf 2}, 46 (1932);
C. Zener,
Non-Adiabatic Crossing of Energy Levels,
Proc. R. Soc. Lond. A {\bf 137}, 696 (1932).

\bibitem{Sphase} St{\"u}ckelberg E. C. G. ,
Theory of inelastic collisions between atoms,
{\it Helv. Phys. Acta} {\bf 5}, 369 (1932).

\bibitem{SA-2-e1600578} Zhou F. , Yan L. L. , Gong S. J. , Ma Z. H. , He J. Z. , Xiong T. P. , Chen L. , Yang W. L. , Feng M. , \& Vedral V. ,
Verifying Heisenberg's error-disturbance relation using a single trapped ion,
{\it Sci. Adv.} {\bf 2}, e1600578 (2016).

\bibitem{njp-19-063032} Xiong T. P. , Yan L. L. , Zhou F. , Rehan K. , Liang D. F. , Chen L. , Yang W. L. , Ma Z. H. , Feng M. , \& Vedral V. ,
Experimental verification of a Jarzynski-related information-theoretic equality using a single trapped ion,
{\it Phys. Rev. Lett.} {\bf 120}, 010601 (2018).

\bibitem{PRR-2-033082} Zhang J. W. , Rehan K. , Li M. , Li J. C. , Chen L. , Su S. L. , Yan L. L. , Zhou F. , \& Feng M. ,
Single-atom verification of the information-theoretical bound of irreversibility at the quantum level,
{\it Phys. Rev. Research} {\bf 2}, 033082 (2020).



\bibitem{explain1} In this model, we can engineer both the Rabi frequency $\Omega$ and the effective dacay rate $\gamma_{\mathrm{eff}}=\tilde{\Omega}^{2}/\Gamma$ under the condition of $\Omega\ll\tilde{\Omega}$ \cite{PRR-2-033082, NC-2022}. With this level of controllability, we can fully tune this two-level system and perform parametric loops that encircle or do not encircle the LEP.

\bibitem{SM} See Supplementary Materials.


\bibitem{pre-76-031105} Quan, H. T., Liu, Y. X., Sun, C. P. \& Nori, F.,
Quantum thermodynamic cycles and quantum heat engine.
{\it Phys. Rev. E} {\bf 76}, 031105 (2007).

\bibitem{pr-697-1} Benenti, G., Casati, G., Saito, K. \& Whitney, R. S.
Fundamental aspects of steady-state conversion of heat to work at the nanoscale.
{\it Phys. Rep.}{\bf 694}, 1 (2017).


\bibitem{pra-103-l020201} Wang, W. C. et al. Observation of PT -symmetric quantum coherence in a single-ion system. Phys. Rev. A 103, L020201 (2021).

\bibitem{prl-118-033901} Cao Q. T. , Wang H. , Dong C. H. , Jing H. , Liu R. S. , Chen X. , Ge L. , Gong Q. , \& Xiao Y. F. ,
Experimental Demonstration of Spontaneous Chirality in a Nonlinear Microresonator,
{\it Phys. Rev. Lett.} {\bf 118}, 033901 (2017).

\bibitem{prl-127-140504} Chen, W., Abbasi, M., Joglekar, Y. N. \& Murch, K. W.
Quantum Jumps in the Non-Hermitian Dynamics of a Superconducting Qubit.
Experimental Demonstration of Spontaneous Chirality in a Nonlinear Microresonator,
{\it Phys. Rev. Lett.} {\bf 127}, 140504 (2021).


\end{thebibliography}

\clearpage

%\addtocounter{figure}{-3}
\section{\centerline{Figures in the maintext}}
\renewcommand{\baselinestretch}{1.5}
\renewcommand\figurename{Maintext Fig.}
\renewcommand\tablename{TABLE S.}
\renewcommand{\andname}{\ignorespaces}
\setcounter{figure}{0}
\setcounter{table}{0}
\begin{figure}[htbp]
\centering
\includegraphics[width=16 cm]{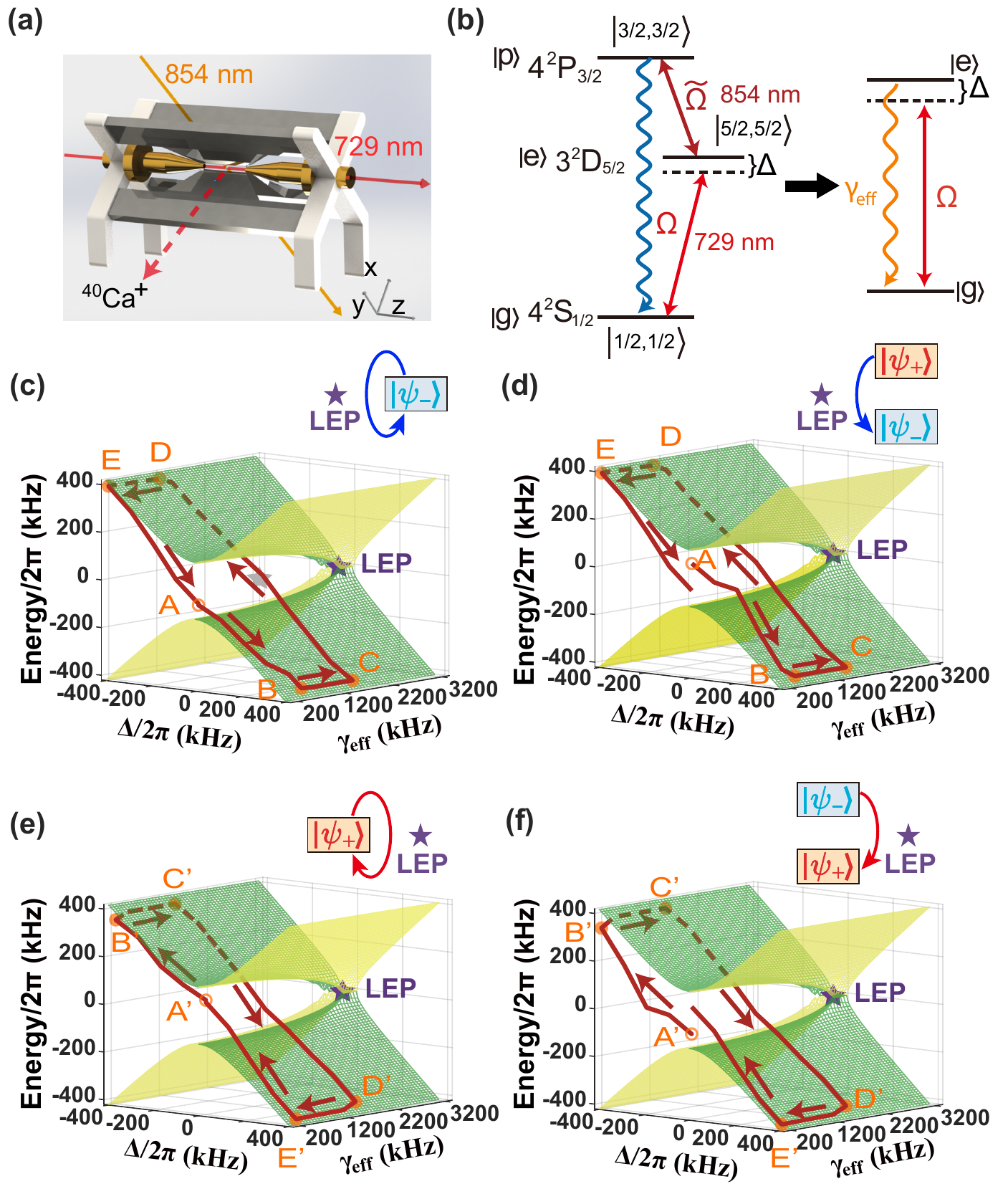}
\caption{Parametric loops in the vicinity of an LEP in the parameter space of a single trapped ion quantum heat engine. (a) The linear Paul trap. (b) Level scheme of the ion, where the solid arrows represent the transitions with Rabi frequencies $\Omega$ and $\tilde{\Omega}$ driven by 729 nm and 854 nm lasers, respectively, and $\Delta$ is the detuning between the energy level and 729 nm laser. The wavy arrow denotes the spontaneous emission with decay rate $\Gamma$. This three-level model can be simplified to an effective two-level system with tunable drive and decay. (c-f) Trajectories without encircling the LEP on the Riemann sheets when completing a clockwise or counterclockwise encirclement in $\Delta\mbox{-}\gamma_{\mathrm{eff}}$ parametric space starting from $|\psi_{+}\rangle$ or $|\psi_{-}\rangle$, where the parameters take the values $\Delta_{\mathrm{min}}/2\pi=-400$ kHz, $\Delta_{\mathrm{max}}/2\pi=400$ kHz, $\gamma_{\mathrm{min}}\approx100$ kHz, $\gamma_{\mathrm{max}}\approx1.45$ MHz. The empty circles represent the starting points. Five orange corner points A, B, C, D, E (A', B', C', D', E') are labeled for convenience of description in the text. The LEPs are labeled by stars.} \label{Fig1}
\end{figure}

\clearpage
\begin{figure*}[htbp]
\centering
\includegraphics[width=17 cm]{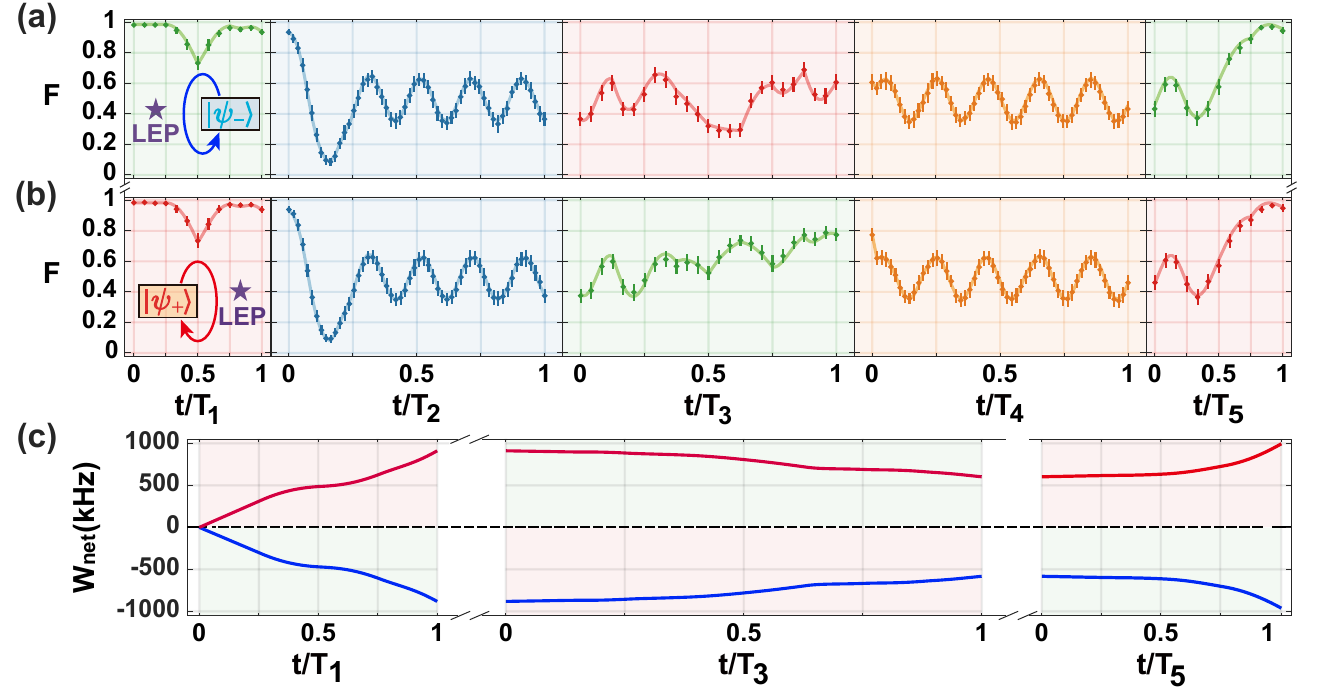}
\caption{Closed loops that do not exhibit mode conversion. (a) Counterclockwise loop starting from $|\psi_{-}\rangle$. (b) Clockwise loop starting from $|\psi_{+}\rangle$. Evolution of the system's state along the trajectory of the closed loop is characterized by the fidelity $\langle\psi_{-}|\rho(t)|\psi_{-}\rangle$ in (a) and the fidelity $\langle\psi_{+}|\rho(t)|\psi_{+}\rangle$ in (b). The circles and error bars denote, respectively, the average and standard deviations of 10000 measurements. The solid curves connecting the dots are obtained by simulating master equations. (c) Blue and red solid curves represent, respectively, the time evolution of the net work in the counterclockwise and clockwise loops, where only three of the strokes associated with doing work are depicted.
%Regions with different colors represent, respectively, different strokes of the heat engine cycle and refrigerator cycles, with the orange and blue corresponding to isochoric heating and isochoric cooling strokes respectively and the red and green denoting iso-decay expansion and iso-decay compression strokes, respectively.
Durations of the five strokes are $T_{1}=T_{5}=6$ $\mu$s, $T_{3}=12$ $\mu$s, and $T_{2}=T_{4}=150$ $\mu$s. Other parameters are $\Omega/2\pi=$ 120 kHz, $\Delta_{\mathrm{min}}/2\pi=-400$ kHz, $\Delta_{\mathrm{max}}/2\pi=400$ kHz, $\gamma_{\mathrm{min}}\approx0$ kHz, and $\gamma_{\mathrm{max}}\approx1.45$ MHz.} \label{Fig2}
\end{figure*}

\clearpage
\begin{figure*}[htbp]
\centering
\includegraphics[width=17cm]{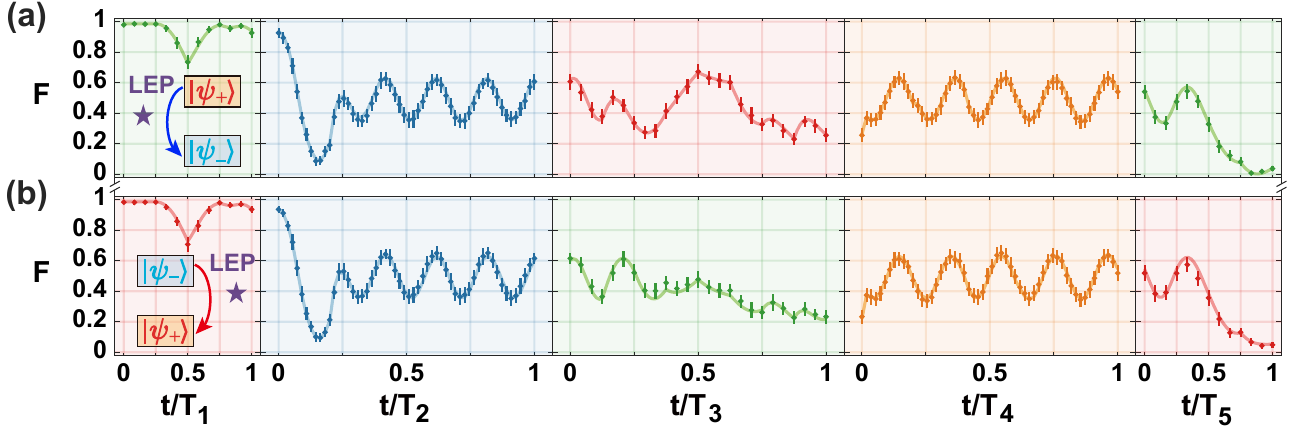}
\caption{Closed loops that lead to mode conversion. (a) Counterclockwise loop starting at $|\psi_{+}\rangle$ and ending at $|\psi_{-}\rangle$. (b) Clockwise loop starting at $|\psi_{-}\rangle$ and ending at $|\psi_{+}\rangle$. Evolution of the system's state along the closed loop trajectory is characterized by the fidelity $\langle\psi_{+}|\rho(t)|\psi_{+}\rangle$ in (a) and the fidelity $\langle\psi_{-}|\rho(t)|\psi_{-}\rangle$ in (b). The circles and error bars respectively denote the average and standard deviations of 10000 measurements. The solid curves are obtained by simulating master equations.
%Regions with different colors represent different strokes respectively, with the orange and blue corresponding to isochoric heating
%and isochoric cooling strokes, respectively, and the red and green denoting iso-decay expansion and iso-decay compression strokes, respectively.
Durations of the five strokes are $T_{1}=T_{5}=6$ $\mu$s, $T_{3}=12$ $\mu$s, and $T_{2}=T_{4}=150$ $\mu$s. Other parameters are $\Omega/2\pi=$ 120 kHz, $\Delta_{\mathrm{min}}/2\pi=-400$ kHz, $\Delta_{\mathrm{max}}/2\pi=400$ kHz, $\gamma_{\mathrm{min}}\approx0$ kHz, and $\gamma_{\mathrm{max}}\approx1.45$ MHz.} \label{Fig3}
\end{figure*}

\newpage

\section{\centerline{Supplementary Materials}}
\renewcommand{\baselinestretch}{1.5}
\renewcommand\figurename{Fig.S.}
\renewcommand\tablename{TABLE S.}
\renewcommand{\andname}{\ignorespaces}
\setcounter{figure}{0}
\setcounter{table}{0}

We present details about the dynamics and chirality of quantum heat engine (QHE) and quantum refrigerator (QR). We also discuss the encirclement within single Riemann sheets, i.e., in half of the encirclement area as considered in the main text.

\subsection*{I. order parameter with respect to Liouvillian exceptional point}
Our system is governed by the Lindblad master equation
\begin{equation}
\dot{\rho}(t)=-i[H_{\mathrm{eff}},\rho ]+\frac{\gamma _{\mathrm{eff}}}{2}%
(2\sigma _{-}\rho \sigma _{+}-\sigma _{+}\sigma _{-}\rho -\rho \sigma
_{+}\sigma _{-})=\mathcal{L}\rho ,  %\nonumber
\label{201}
\end{equation}
where $\mathcal{L}$ is called Liouvillian superoperator and $\rho$ is the density operator of the system. Here the density matrix $\rho$ is expressed by
\begin{math}
\left (
\begin{array}{ccc}
\rho_{ee}&\rho_{eg}\\\rho_{ge}&\rho_{gg}
\end{array}
\right )
\end{math}, with the raising and lowering operators defined as $\sigma_{+}=\left (\begin{array}{ccc} 0 & 1 \\ 0 & 0 \end{array}
\right )$ and $\sigma_{-}=\left (\begin{array}{ccc} 0 & 0 \\ 1 & 0 \end{array}\right )$, respectively. According to the corresponding equations in the main text, we can write $\mathcal{L}$ as
\begin{equation}
\mathcal{L}=\left(
\begin{array}{cccc}
-\gamma _{\mathrm{eff}} & i\Omega /2 & -i\Omega /2 & 0 \\
i\Omega /2 & -(\gamma _{\mathrm{eff}}/2+i\Delta ) & 0 & -i\Omega /2 \\
-i\Omega /2 & 0 & -(\gamma _{\mathrm{eff}}/2-i\Delta ) & i\Omega /2 \\
\gamma _{\mathrm{eff}} & -i\Omega /2 & i\Omega /2 & 0%
\end{array}
\right).
\label{L202}
\end{equation}
By setting $\Delta =0$, we find the eigenvalues of $\mathcal{L}$ as $\lambda_{1}=0$, $\lambda _{2}=-\gamma_{\mathrm{eff}}/2$, $\lambda_{3}=\frac{1}{4}
(-3\gamma_{\mathrm{eff}}-\sqrt{\gamma_{\mathrm{eff}}^{2}-16\Omega^{2}})$, and $\lambda_{4}=\frac{1}{4}(-3\gamma_{\mathrm{eff}}+\sqrt{\gamma_{\mathrm{eff}}^{2}-16\Omega^{2}})$. It is evident that the eigenvalues $\lambda_{3}$ and $\lambda_{4}$ becomes degenerate at $\gamma_\mathrm{eff}=4\Omega$, which corresponds to the Liouvillian exceptional point (LEP). For $\gamma_{\mathrm{eff}}>4\Omega$, we have real $\lambda_{3}$ and $\lambda_{4}$ with a splitting of $|\lambda_{4}-\lambda_{3}|=\xi/2$ with $\xi=\sqrt{\gamma_{\mathrm{eff}}^2-16\Omega^2}$. For $\gamma_{\mathrm{eff}}<4\Omega$, on the other hand, $\lambda_3$ and $\lambda_4$ form a complex conjugate pair with a splitting of $|\xi|/2$ in their imaginary parts.

%We consider $Im[\lambda]$ as the order parameter of this phase transition, where $\lambda=(-3\gamma_\mathrm{eff}+\sqrt{\gamma_\mathrm{eff}^2-16\Omega^2})/4$, and $Im[\lambda]$  is the real part of the eigenvalues of the system, see Methods of the main text.
\begin{figure}[tbph]
\includegraphics[width=10cm]{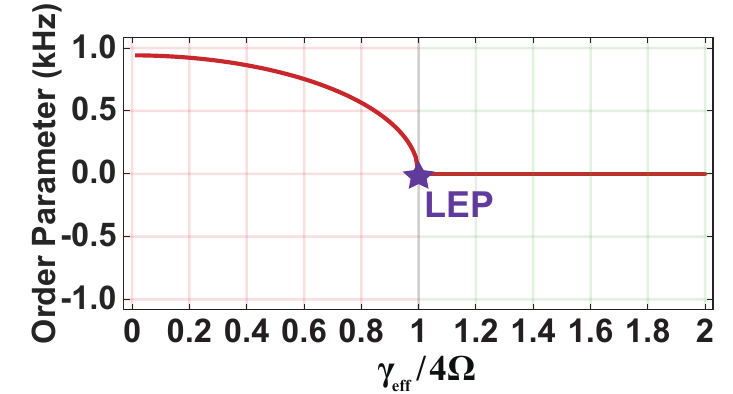}
\caption{Topological phase transition with respect to the Liouvillian exceptional point (LEP) that is determined by the decay rate $\protect\gamma_{\mathrm{eff}}$ and the drive strength $\Omega$. }\label{FigS1}
\end{figure}

The topological phase transition occurs at the LEP, as shown in Fig.S. \ref{FigS1}. Since the LEP is excluded in our displayed encirclements, no topological phase transition occurs in our experiment. Nevertheless, the influence from the LEP is witnessed in our experiment. As shown in the following sections, we have observed experimentally the strong relevance of the chirality to the LEP and the associated Riemann sheets.

{\color{black}
In contrast, with the quantum jumps ignored, the above master equation
(\ref{201}) is reduced to
\begin{equation}
\dot{\rho}(t)=-i[H_{\mathrm{eff}},\rho ]-\frac{\gamma _{\mathrm{eff}}}{2}%
(\sigma _{+}\sigma _{-}\rho +\rho \sigma _{+}\sigma _{-})=-i(H_{\mathrm{NH}%
}\rho -\rho H_{\mathrm{NH}}^{\dag })=\mathcal{L}_{H}\rho ,  %\nonumber
\label{2010}
\end{equation}%
where $H_{\mathrm{NH}}=H_{\mathrm{eff}}-i\frac{\gamma _{\mathrm{eff}}}{2}%
\sigma _{+}\sigma _{-}$ is the non-Hermitian Hamiltonian and
\begin{equation}
\mathcal{L}_{H}=\left(
\begin{array}{cccc}
-\gamma _{\mathrm{eff}} & i\Omega /2 & -i\Omega /2 & 0 \\
i\Omega /2 & -(\gamma _{\mathrm{eff}}/2+i\Delta ) & 0 & -i\Omega /2 \\
-i\Omega /2 & 0 & -(\gamma _{\mathrm{eff}}/2-i\Delta ) & i\Omega /2 \\
0 & -i\Omega /2 & i\Omega /2 & 0%
\end{array}%
\right) .  %\nonumber
\label{H202}
\end{equation}%
is the Liouvillian operator without quantum jumps. Both $H_{\mathrm{NH}}$
and $\mathcal{L}_{H}$\ have been utilized to experimentally verify the phase
transition of EP for $\Delta =0$ \cite{pra-103-l020201}. The above
derivations illustrate, without involving the quantum jump item $\sigma_{-}\rho \sigma
_{+}$, the LEP returns to the Hamiltonian EP (HEP). In other words,
the HEP cannot describe the entire dynamics of the open quantum system.

We find the eigenvalues of $\mathcal{L}_{H}$ as $\tilde{%
\lambda}_{1}=\tilde{\lambda}_{2}=-\gamma _{\mathrm{eff}}/2$, $\tilde{\lambda}%
_{3}=(-\gamma _{\mathrm{eff}}-\sqrt{\gamma _{\mathrm{eff}}^{2}-4\Omega ^{2}}%
)/2$ and $\tilde{\lambda}_{4}=(-\gamma _{\mathrm{eff}}+\sqrt{\gamma _{%
\mathrm{eff}}^{2}-4\Omega ^{2}})/2$. The eigenvalues $\tilde{\lambda}_{3}$
and $\tilde{\lambda}_{4}$ become degenerate at $\gamma _{%
\mathrm{eff}}=2\Omega $, corresponding to the HEP. Real $\tilde{%
\lambda}_{3}$ and $\tilde{\lambda}_{4}$ present a splitting of $|\tilde{\lambda%
}_{3}-\tilde{\lambda}_{4}|=\sqrt{\gamma _{\mathrm{eff}}^{2}-4\Omega^{2}}$ when $\gamma_{\mathrm{eff}}>2\Omega $, while $\tilde{\lambda}_{3}$ and $%
\tilde{\lambda}_{4}$ become a complex conjugate pair with a splitting of $\sqrt{\gamma_{\mathrm{eff}}^{2}-4\Omega^{2}}$ in their imaginary parts.

Considering the experimental parameters and the presence of quantum jumps,
we conclude that the EP observed in our system is an LEP rather than a HEP.
LEP and HEP have been observed in previous experiments \cite{prl-127-140504} and \cite{NP-15-1232}, respectively.
These experiments illustrate that, when the decay rate is fixed, the coupling strength for the LEP \cite{prl-127-140504} is much smaller
than that for the HEP \cite{NP-15-1232}, and this difference results from quantum jumps. Additionally, in the case of $\gamma_{\mathrm{eff}}=0$, the eigenenergies of the LEP and HEP are identical, and we can obtain $\lambda_{1}=\lambda_{2}=\tilde{\lambda}_{2}=\tilde{\lambda}_{1}=0$, $\lambda_{3}=\tilde{\lambda}_{3}=i\Omega$, and $\lambda_{4}=\tilde{\lambda}_{4}=-i\Omega$ for the identical Lindblad operators (\ref{L202}) and (\ref{H202}). In this case, the dynamics of the LEP and HEP share identical results with the same initial states.
}

{\color{black}Besides, in our experiments, the Rabi frequency is $\Omega/2\pi=120$ kHz. Based on other experimental parameters, we have theoretically calculated the decay rate for the LEP as $\gamma_{\mathrm{LEP}}=4\Omega\approx3.0$ MHz, and for the HEP as $\gamma_{\mathrm{HEP}}=2\Omega\approx1.5$ MHz. None of these decay rates is smaller than the maximum decay rate $\gamma_{\mathrm{max}}\approx1.45$ MHz in our experiment. As a result, although there is no HEP in our experiments, we can say, neither the HEPs nor the LEPs are encircled in our experimental loops (see Fig.S. \ref{smplus}).
\begin{figure}[tbph]
\includegraphics[width=16cm]{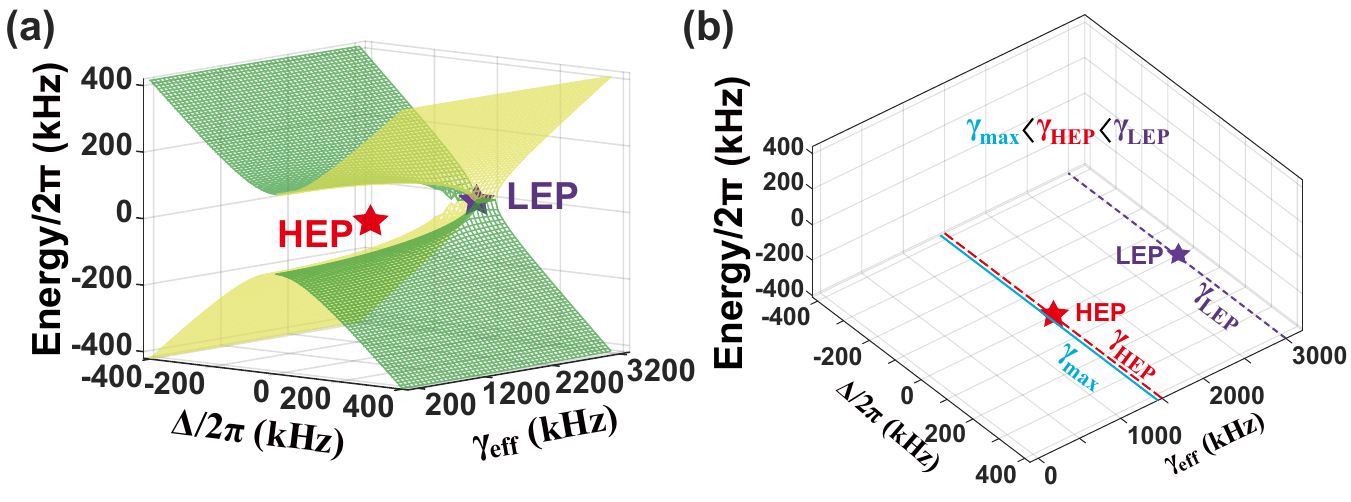}
\caption{(a) Positions of the LEP and HEP in the parameter space for the eigenenergy Riemann surface. It shows, on the eigenenergy Riemann surface, the EPs presented correspond to the LEPs rather than the HEPs.
(b) Projected decay rates ($\gamma_{\mathrm{max}}$, $\gamma_{\mathrm{HEP}}$, and $\gamma_{\mathrm{LEP}}$) in the $\Delta-\gamma_{\mathrm{eff}}$ plane with the Rabi frequency of $\Omega = 120$ kHz.}\label{smplus}
\end{figure}
}

\subsection* {II. Thermodynamic quantities}
We present here the definitions of the thermodynamic quantities, that are the net work $W$ and the efficiency $\eta$. For our trapped-ion system, we define the internal energy of the {\color{black}quantum} heat engine as $U=\mathrm{tr}(\rho H)$, where $\rho$ and $H$ are the density matrix and Hamiltonian of the {\color{black}quantum} system, respectively. In our experiments, the
Hamiltonian of the working substance is given by $H=\Delta\left\vert e\right\rangle\left\langle e\right\vert$. The coupling strength $\Omega$ and the effective dissipation rate $\gamma
_{\mathrm{eff}}$ are employed to tune the system such that the QHE cycle and the QR cycle are performed, as explained in the main text.

In classical thermodynamics, the first law of thermodynamics is expressed as $dU=dW+dQ$. In contrast, in quantum thermodynamics, it is expressed as $dU=$
$d(\mathrm{tr}(\rho H))=\mathrm{tr}(\rho dH)+\mathrm{tr}(Hd\rho )$, where the work and heat in the differential form are defined as $dW=\rho dH$ and $
dQ=Hd\rho$, respectively. Then the work $W_{\mathrm{in}}$ done by the environment and the one $W_{\mathrm{out}}$ by the working substance are given
by $W_{\mathrm{in}}=-\sum\limits_{i}\rho _{i}dH_{i}^{\prime}$ (for $dH_{i}^{\prime }>0$) and $W_{\mathrm{out}}=-\sum\limits_{i}\rho
_{i}dH_{i}^{\prime }$ (for $dH_{i}^{\prime }<0$), respectively. As discussed in the main text, $W_{_{\mathrm{in}}}$ and $W_{\mathrm{%
out}}$ are calculated, respectively, in the iso-decay compression and iso-decay expansion strokes of our trapped-ion QHE and QR cycles. Thus, the net work acquired can be
written as
\begin{equation}
\begin{array}{ccc}
W_{\mathrm{net}} & = & W_{_{\mathrm{in}}}+W_{\mathrm{out}}=-\sum\limits_{i}
\rho _{i}dH_{i}^{\prime },
\end{array}   \nonumber
\label{401}
\end{equation}%
which is used to calculate the net work in the main text Figure 2.

\subsection*{III. Dynamics of quantum heat engine and quantum refrigerator cycles}
\begin{figure}[tbph]
\includegraphics[width=16cm]{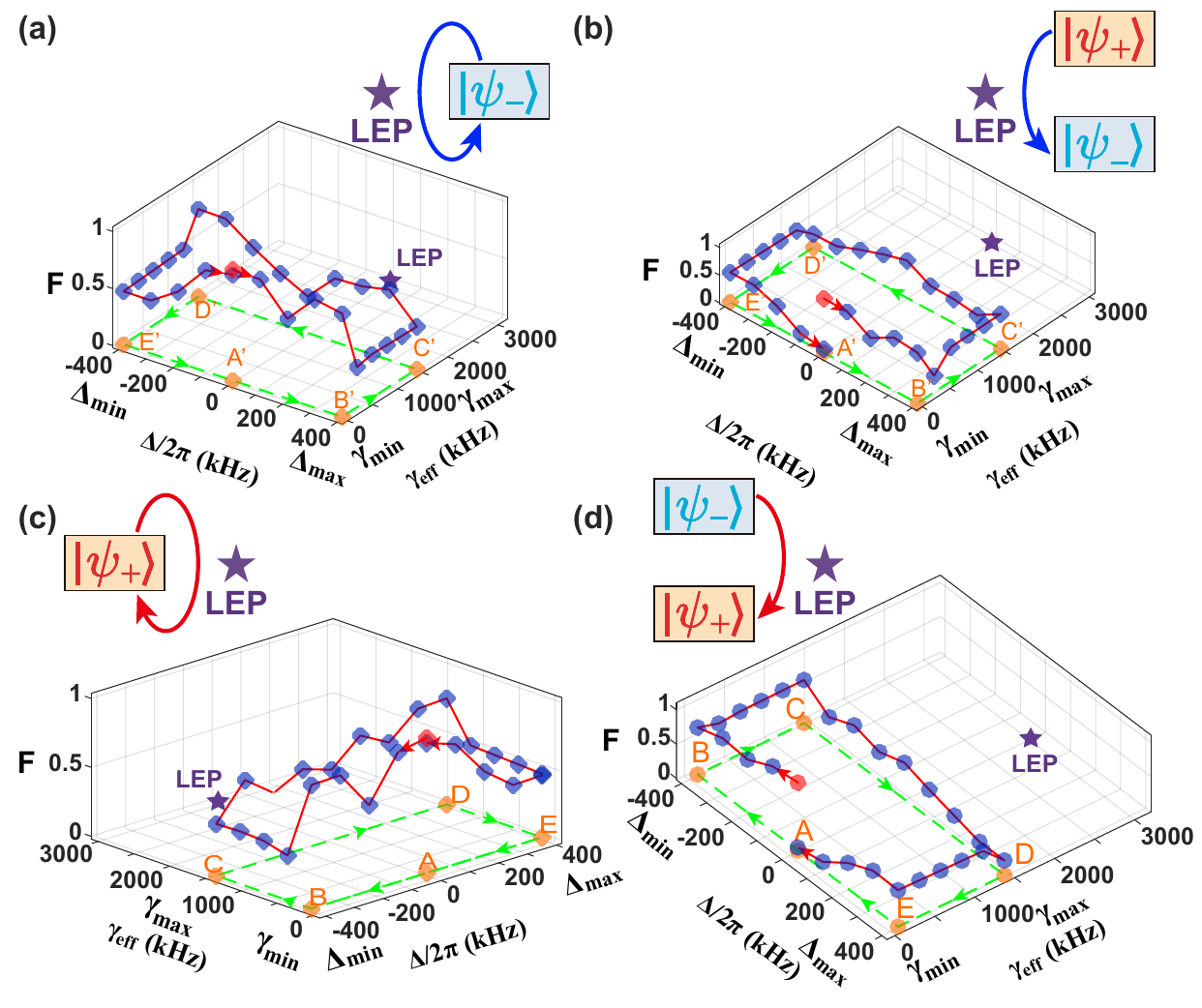}
\caption{Variation of fidelity with respect to the detuning and the effective decay rate $\gamma_{\mathrm{eff}}$, where we consider the fidelity $\langle\psi_{+}|\rho(t)|\psi_{+}\rangle$ or $\langle\psi_{-}|\rho(t)|\psi_{-}\rangle$ when completing a clockwise or a counterclockwise encirclement, employing
$\Delta_{\mathrm{min}}/2\pi=-400$ kHz, $\Delta_{\mathrm{max}}/2\pi=400$ kHz, $\gamma_{\mathrm{min}}\approx0$ kHz, and $\gamma_{\mathrm{max}}\approx1.45$ MHz. The red dots represent the starting points and the red lines show the evolution trends, instead of the real evolution trajectories, with the red arrows denoting the encircling directions. The green dashed curves are the projection of the solid red curves on the bottom plane for guiding eyes. Five orange corner points A, B, C, D, E (A', B', C', D', E') are labeled for convenience of description in the text. The LEPs are labeled by the purple stars.}
\label{FigS2}
\end{figure}

We execute the clockwise and counterclockwise encirclements without encircling the LEPs as designed in the main text by elaborately controlling
the 854-nm laser (for tuning $\gamma_{\mathrm{eff}}$) and the 729-nm laser (for tuning $\Delta$) with $\Omega/2\pi\approx120$ kHz in the whole process. The dynamical processes of the four encirclements are demonstrated by observing the evolution of the fidelity $\langle\psi_{+}|\rho(t)|\psi_{+}\rangle$ or $\langle\psi_{-}|\rho(t)|\psi_{-}\rangle$ as shown in Fig. S~\ref{FigS2}(a-d).

For the clockwise QHE cycle in Fig.S.~\ref{FigS2}(c), we first prepare the state of the system in $|g\rangle$ and then apply a $\pi/2$ pulse to prepare the initial state $|\psi_{+}\rangle=(|e\rangle+|g\rangle)/\sqrt{2}$, corresponding to $\rho_{A}=$\begin{math}\left(\begin{smallmatrix}0.4925&0.4925 \\ 0.4925&0.4925\end{smallmatrix}\right)\end{math} experimentally (due to the imperfect ground state preparation and subsequent $\pi/2$ pulse operation), and thus the initial fidelity is $\langle\psi_{+}|\rho_{A}|\psi_{+}\rangle=0.985$. Then we execute the first iso-decay expansion process by decreasing the detuning $\Delta$ from $\Delta=0$ kHz to $\Delta_{\mathrm{min}}=-2\pi\times400$ kHz while $\gamma_{\mathrm{eff}}\approx0$ kHz remains unchanged with $T_{1}=6$ $\mu$s. The system evolves to $\rho_{B}=$\begin{math}\left(\begin{smallmatrix}0.668&0.444 - 0.122i \\ 0.444 + 0.122i&0.317\end{smallmatrix}\right)\end{math} as numerically calculated. Then we carry out the isochoric cooling process by increasing the decay rate $\gamma_{\mathrm{eff}}$ from $\gamma_{\mathrm{min}}\approx0$ kHz to $\gamma _{\mathrm{max}}\approx1.43$ MHz while $\Delta=\Delta_{\mathrm{min}}$ is fixed. Experimentally, we separate the evolution into five steps with the increasing decay rate, i.e. $T_{2-1}=30$ $\mu$s, $\gamma_{\mathrm{eff}}\approx0$ kHz; $T_{2-2}=30$ $\mu$s, $\gamma_{\mathrm{eff}}\approx286$ kHz; $T_{2-3}=30$ $\mu$s, $\gamma_{\mathrm{eff}}\approx572$ kHz; $T_{2-4}=30$ $\mu$s, $\gamma_{\mathrm{eff}}\approx858$ kHz; $T_{2-5}=30$ $\mu$s, $\gamma_{\mathrm{eff}}\approx1.14$ MHz, the system evolves to $\rho_{C}=$\begin{math}\left(\begin{smallmatrix}0.021&-0.107 - 0.089i \\ -0.107 + 0.089i&0.964\end{smallmatrix}\right)\end{math} after this process due to the large decay and detuning. The next stroke is the iso-decay compression with the detuning tuned from $\Delta_{\rm{min}}$ to $\Delta_{\mathrm{max}}$ and the constant value of $\gamma_{\mathrm{eff}}=\gamma_{\mathrm{max}}$, the duration of this stroke is $T_{3}=12$ $\mu$s, and thus the system evolves to $\rho_{D}=$\begin{math}\left(\begin{smallmatrix}0.104&0.284 - 0.008i \\ 0.284 + 0.008i&0.88\end{smallmatrix}\right)\end{math}. The fourth step is the isochoric heating process with the decay rate $\gamma_{\mathrm{eff}}$ tuned from $\gamma_{\mathrm{max}}\approx1.43$ MHz to $\gamma_{\mathrm{min}}\approx0$ kHz when $\Delta=\Delta_{\mathrm{max}}$ is fixed. During this process, our experimental operation is to separate the evolution into five steps along with the decay, i.e., $T_{4-1}=30$ $\mu$s, $\gamma_{\mathrm{eff}}\approx1.43$ MHz; $T_{4-2}=30$ $\mu$s, $\gamma_{\mathrm{eff}}\approx1.14$ MHz; $T_{4-3}=30$ $\mu$s, $\gamma_{\mathrm{eff}}\approx858$ kHz; $T_{4-4}=30$ $\mu$s, $\gamma_{\mathrm{eff}}\approx572$ kHz; $T_{4-5}=30$ $\mu$s, $\gamma_{\mathrm{eff}}\approx286$ kHz, the system evolves to a steady state $\rho_{E}=$\begin{math}\left(\begin{smallmatrix}0.021&-0.036 - 0.137i \\ -0.036 +0.137i&0.964\end{smallmatrix}\right)\end{math} after this stroke with negligible non-diagonal elements due to the large decay and the long duration time. The final step is the iso-decay expansion stroke process by decreasing the detuning from $\Delta=2\pi\times400$ kHz to $\Delta=0$ kHz while $\gamma_{\mathrm{eff}}=\gamma_{\mathrm{min}}\approx0$ kHz remains unchanged with $T_{5}=6$ $\mu$s, and thus the system experiences a Landau-Zener transition and accumulate a St{\"u}ckelberg phase, reaching the final state \begin{math}\left(\begin{smallmatrix}0.651&0.464-0.044i \\ 0.464+0.044i&0.334\end{smallmatrix}\right)\end{math}. The calculated final fidelity is $\langle\psi_{+}|\rho(t)|\psi_{+}\rangle=0.957$, indicating that the system finally returns to the initial state after the encirclement.

By contrast, for the counterclockwise QR cycle in Fig.S. \ref{FigS2}(a), we prepare the initial state $|\psi_{0}\rangle=|\psi_{-}\rangle=(|e\rangle-|g\rangle)/\sqrt{2}$, corresponding to $\rho_{A'}=$\begin{math}\left(\begin{smallmatrix}0.4925&-0.4925 \\ -0.4925&0.4925\end{smallmatrix}\right)\end{math} experimentally, and the initial fidelity is $\langle\psi_{-}|\rho(A')|\psi_{-}\rangle=0.985$. We first execute the iso-decay compression stroke by increasing the detuning $\Delta$ from $\Delta=0$ kHz to $\Delta_{\rm{max}}=2\pi\times400$ kHz while $\gamma_{\mathrm{eff}}\approx0$ kHz remains unchanged with $T_{1}=6$ $\mu$s, and thus the system evolves to $\rho_{B'}=$\begin{math}\left(\begin{smallmatrix}0.668&-0.444+0.122i \\ -0.444-0.122i&0.317\end{smallmatrix}\right)\end{math}. Then we carry out the isochoric cooling stroke by increasing the decay rate $\gamma_{\mathrm{eff}}$ from $\gamma_{\mathrm{min}}\approx0$ kHz to $\gamma_{\mathrm{max}}=1.43$ MHz while $\Delta=\Delta_{\mathrm{max}}$ is fixed. Experimentally, we separate the evolution into five steps with the increasing decay rate, i.e. $T_{2-1}=30$ $\mu$s, $\gamma_{\mathrm{eff}}\approx0$ kHz; $T_{2-2}=30$ $\mu$s, $\gamma_{\mathrm{eff}}\approx286$ kHz; $T_{2-3}=30$ $\mu$s, $\gamma_{\mathrm{eff}}\approx572$ kHz; $T_{2-4}=30$ $\mu$s, $\gamma_{\mathrm{eff}}\approx858$ kHz; $T_{2-5}=30$ $\mu$s, $\gamma_{\mathrm{eff}}\approx1.14$ MHz, and thus the system evolves to $\rho_{C'}=$\begin{math}\left(\begin{smallmatrix}0.021&0.128-0.055i \\ -0.128+0.055i&0.964\end{smallmatrix}\right)\end{math} after this stroke. The next stroke is the iso-decay expansion with the detuning tuned from $\Delta_{\mathrm{max}}$ to $\Delta_{\mathrm{min}}$ with the constant value of $\gamma_{\mathrm{eff}}=\gamma_{\mathrm{max}}$ with $T_{3}=12$ $\mu$s, and thus the system evolves to $\rho_{D'}=$\begin{math}\left(\begin{smallmatrix}0.098&-0.112-0.26i \\ -0.112+0.26i&0.886\end{smallmatrix}\right)\end{math} after this stroke. The fourth step is the isochoric heating process with decay rate $\gamma_{\mathrm{eff}}$ tuned from $\gamma_{\mathrm{max}}\approx1.43$ MHz to $\gamma_{\mathrm{min}}\approx0$ kHz when $\Delta=\Delta_{{\mathrm{min}}}$ is fixed. During this process, we divide the evolution into five stages of the decay rate, i.e., $T_{4-1}=30$ $\mu$s, $\gamma_{\mathrm{eff}}\approx1.43$ MHz; $T_{4-2}=30$ $\mu$s, $\gamma_{\mathrm{eff}}\approx1.14$ MHz; $T_{4-3}=30$ $\mu$s, $\gamma_{\mathrm{eff}}\approx858$ kHz; $T_{4-4}=30$ $\mu$s, $\gamma_{\mathrm{eff}}\approx572$ kHz; $T_{4-5}=30$ $\mu$s, $\gamma_{\mathrm{eff}}\approx286$ kHz, and thus the system evolves to a steady state $\rho_{E'}=$\begin{math}\left(\begin{smallmatrix}0.021&0.036- 0.137i \\0.037+0.137i&0.964\end{smallmatrix}\right)\end{math} after this stroke with negligible non-diagonal elements. The final step is the iso-decay compression stroke by increasing the detuning from $\Delta=-2\pi\times400$ kHz to $\Delta=0$ kHz while $\gamma_{\mathrm{eff}}=\gamma_{\mathrm{min}}\approx0$ kHz remains unchanged with $T_{5}=6$ $\mu$s. The system experiences a LZS process and accumulates a St{\"u}ckelberg phase contrary to the clockwise encirclements, thus reaching the final state \begin{math}\left(\begin{smallmatrix}0.651&-0.464-0.044i \\ -0.464+0.044i&0.334\end{smallmatrix}\right)\end{math}. The final fidelity is $\langle\psi_{-}|\rho(t)|\psi_{-}\rangle=0.957$, indicating that the system almost evolves to the initial state at the end of the encirclement.

While for the clockwise encirclement depicted in Fig.S. \ref{FigS2}(d), we prepare the state initially in $|\psi_{0}\rangle=|\psi_{-}\rangle=(|e\rangle-|g\rangle)/\sqrt{2}$, corresponding to $\rho_{A}=$\begin{math}\left(\begin{smallmatrix}0.4925&-0.4925 \\ -0.4925&0.4925\end{smallmatrix}\right)\end{math} experimentally, the fidelity is initially $\langle\psi_{-}|\rho(A)|\psi_{-}\rangle=0.985$. We repeat the experimental sequence of the QHE cycle, then the system evolves to a steady state $\rho_{E}=$\begin{math}\left(\begin{smallmatrix}0.021&-0.036 - 0.137i \\ -0.036 + 0.137i&0.964\end{smallmatrix}\right)\end{math} after the fourth stroke (the same as the QHE cycle). After experiencing a LZS in the final stroke, the system evolves to \begin{math}\left(\begin{smallmatrix}0.453&0.445+0.206i \\ 0.445-0.206i&0.532\end{smallmatrix}\right)\end{math}. The final fidelity is $\langle\psi_{-}|\rho(t)|\psi_{-}\rangle=0.0487$, showing the system cannot return to the initial state after the encirclement. For the counterclockwise encirclement depicted in Fig.S. \ref{FigS2}(b), we prepare  the initial state $|\psi_{0}\rangle=|\psi_{+}\rangle=(|e\rangle+|g\rangle)/\sqrt{2}$, corresponding to $\rho_{A'}=$\begin{math}\left(\begin{smallmatrix}0.4925&0.4925 \\ 0.4925&0.4925\end{smallmatrix}\right)\end{math} experimentally, the initial fidelity is $\langle\psi_{+}|\rho(A')|\psi_{+}\rangle=0.985$.  We repeat the experimental sequence of the QR cycle, then the system evolves to a steady state $\rho_{E'}=$\begin{math}\left(\begin{smallmatrix}0.021&0.058 - 0.129i \\ 0.058 + 0.129i&0.964\end{smallmatrix}\right)\end{math} after the fourth stroke (the same as the QR cycle). After the fifth stroke, the system evolves to \begin{math}\left(\begin{smallmatrix}0.667&-0.454-0.075i \\ -0.454+0.075i&0.318\end{smallmatrix}\right)\end{math}. The final fidelity is $\langle\psi_{+}|\rho(t)|\psi_{+}\rangle=0.039$, showing that the system cannot evolve back to the initial state after the encirclement.

As a consequence, we conclude that the chirality depends on both the initial state and the encircling direction, which results in different thermodynamic processes, i.e., either QHE or QR.

\subsection*{IV. State evolution with respect to eigenstates}

\begin{figure}[tbph]
\includegraphics[width=16cm]{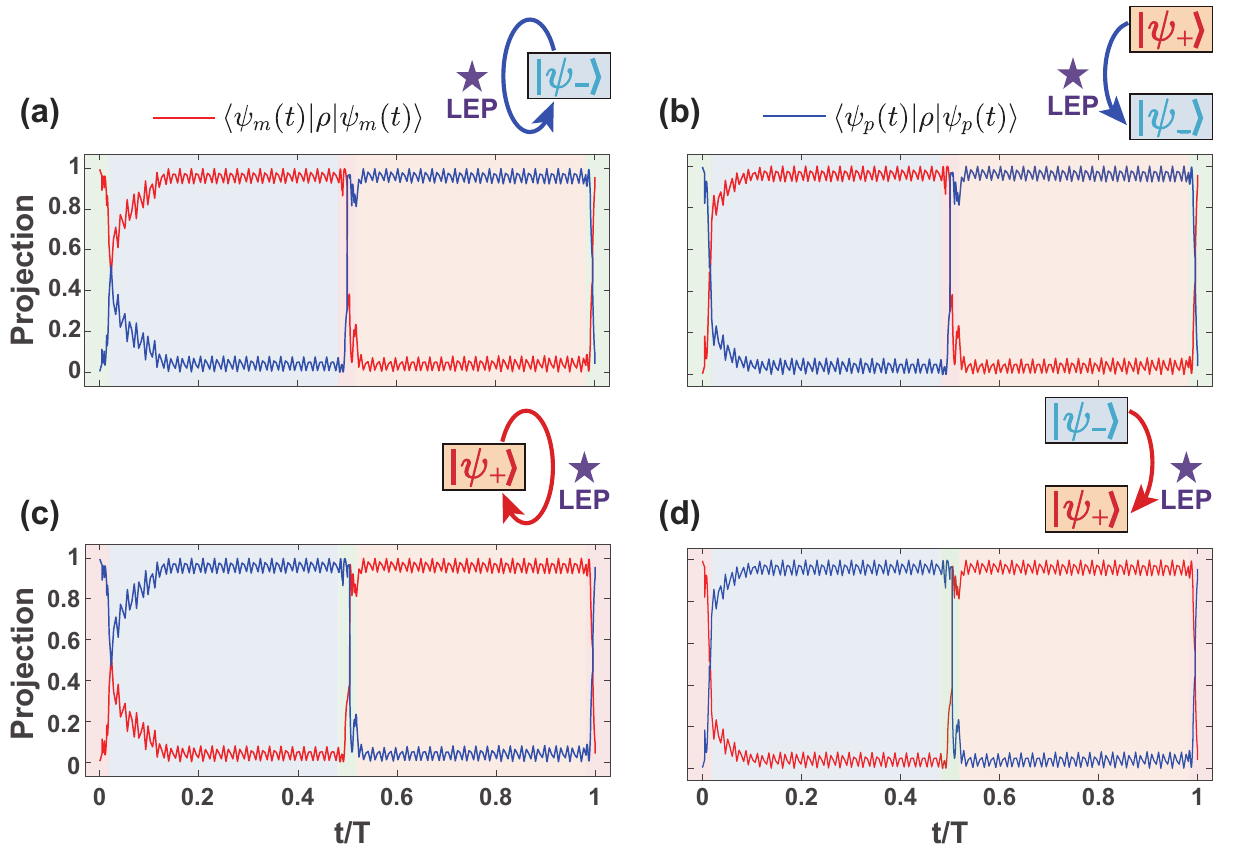}
\caption{State evolution with respect to the eigenstates of the effective Hamiltonian. The evolution of the encircling state for clockwise (counterclockwise) loops starting from $|\psi_{+}\rangle$ ($|\psi_{-}\rangle$) is characterized by the projection $\langle\psi_{m}(t)|\rho|\psi_{m}(t)\rangle$  ($\langle\psi_{p}(t)|\rho|\psi_{p}(t)\rangle$), where $\psi_{p}(t)$ and $\psi_{m}(t)$ correspond, respectively, to the upper and lower branches of the Riemann surface.}
\label{FigS3}
\end{figure}

We numerically calculate the state evolutions with respect to the eigenstates of the time-dependent effective Hamiltonian $H_{\mathrm{eff}}=(\Delta-i\gamma_{\mathrm{eff}})|e\rangle\langle e|+\Omega/2(|e\rangle\langle g|+|g\rangle\langle e|)$. The encirclements plotted in Figs.S. \ref{FigS3}(a)-(d) correspond to the trajectories depicted in Figs. 1(c)-(f) in the main text, respectively.

\subsection{V. Variation of the chirality when approaching LEP}

In this section, we carry out numerical simulation for the fidelity after completing the clockwise or counterclockwise encirclements. Considering the initial states $|\psi_{+}\rangle$ and $|\psi_{-}\rangle$, respectively, we sweep $\gamma_{\mathrm{max}}$ but fix other parameters, monitoring how the population of the final state varies versus $\gamma_{\mathrm{max}}$.
\begin{figure}[tbph]
\includegraphics[width=16cm]{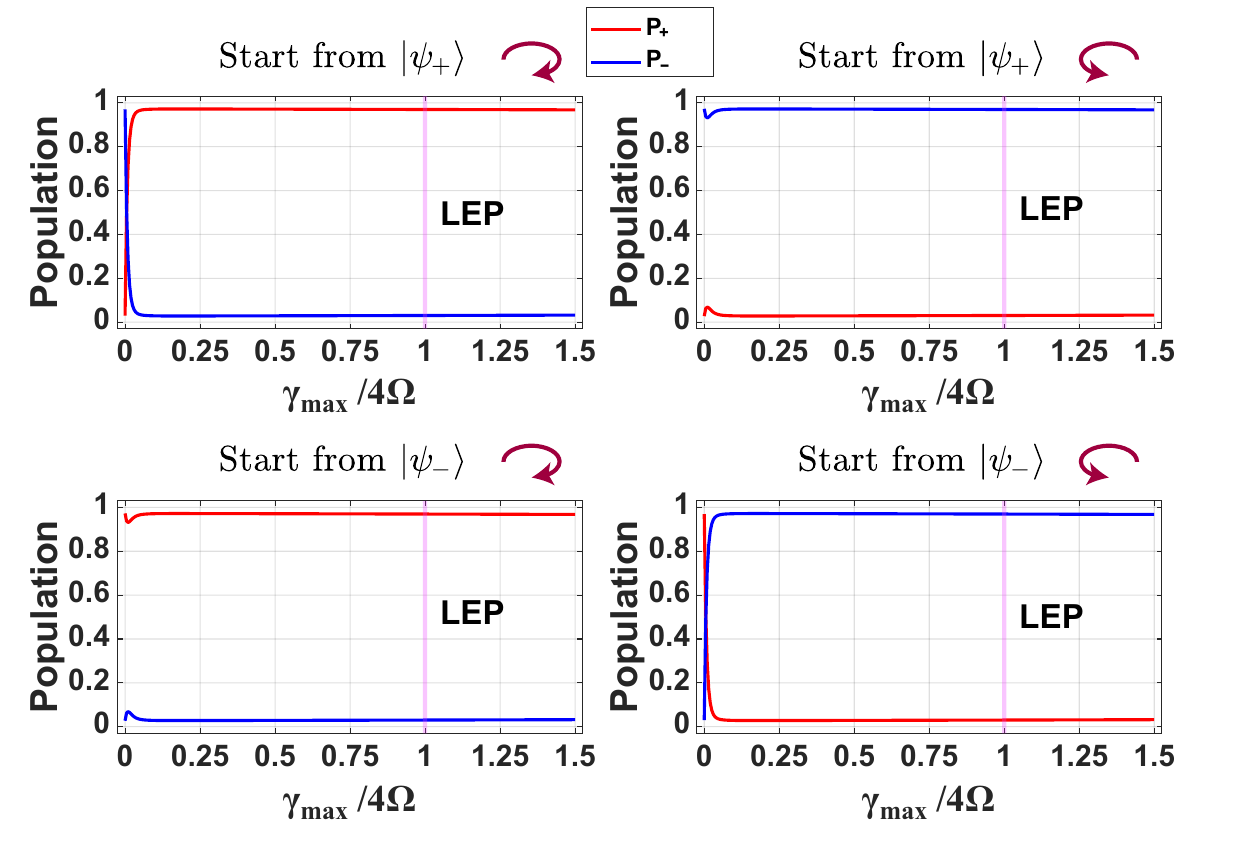}
\caption{Populations of the final states with respect to $\gamma_{\rm{max}}/4\Omega$ in different cases of the initial state and encircling direction. Calculations are made by following parameter values: $\Delta_{\mathrm{min}}/2\pi=-400$ kHz, $\Delta_{\mathrm{max}}/2\pi=400$ kHz, $\gamma_{\mathrm{min}}\approx0$ kHz, $T_{1}=T_{5}=6$ $\mu$s, $T_{2}=T_{4}=50$ $\mu$s, $T_{3}=12$ $\mu$s. Red and blue solid curves represent the populations of $|\psi_{+}\rangle$ and $|\psi_{-}\rangle$, respectively. Pink lines mark the positions of LEPs.}
\label{FigS4}
\end{figure}

We find a transition from no chirality to chirality with the increasing decay rate $\gamma_\mathrm{max}$. When the decay rate $\gamma_\mathrm{max}$ is small enough, the Landau-Zener--St{\"u}ckelberg (LZS) process ensures the final state to be the same as the initial state and no asymmetric mode conversion occurs. As plotted in Fig.S. \ref{FigS4}, we see that both the clockwise and counterclockwise encirclements starting from $|\psi_{+}\rangle$ evolve to the same final state $|\psi_{+}\rangle$, and the system finally reaches $|\psi_{-}\rangle$ when starting from $|\psi_{-}\rangle$. With the increase of the decay rate $\gamma_\mathrm{max}$, however, we see asymmetric mode conversion occurring as $\gamma_\mathrm{max}/4\Omega\ge$ 0.05. This is due to the non-adiabatic transition induced by the enhanced decay. As a result, the chirality appears. Taking the initial state $|\psi_{+}\rangle$ as an example, we find the non-adiabatic transition leading to a steady state after the fourth stroke when encircling along the clockwise direction. Then an LZS process results in the final state $|\psi_{-}\rangle$. When a counterclockwise encirclement is performed, the state evolves back to $|\psi_{+}\rangle$ ultimately. {\color{black} As a result, the decay rate $\gamma_\mathrm{max}$ plays a key role in our experiment. Combined with the Landau-Zener transition, it leads to non-adiabatic transitions, resulting in chiral behavior and asymmetric mode conversion (see the starting points for $\gamma_\mathrm{max}\approx0$ in Fig.S. \ref{FigS4}).}

We consider that it is possible to have an analytical estimate of the critical decay rate for the transition, although writing a specifically analytical relation of the chirality transition with the decay rate is challenging. The simplest way for such an estimate is to consider the shortest time required for accomplishing a chirality transition. To this end, we assume that the dissipation occurs only in the iso-decay process of the 3rd stroke with the decay rate $\gamma_\mathrm{max}$. Then the shortest time for the system to achieve its steady state is
 $T_\mathrm{min}=1/\gamma_\mathrm{max}$.  If the evolution time of the 3rd stroke is $T=12$ $\mu$s, the minimum decay rate for the chirality can be simply estimated as $\gamma_\mathrm{min}/4\Omega=(1/T)/4\Omega\approx0.0276$. This complies with the numerical result in Fig.S.~\ref{FigS4}.

\subsection*{VI. Nonreciprocal chirality}

\begin{figure}[tbph]
\includegraphics[width=16cm]{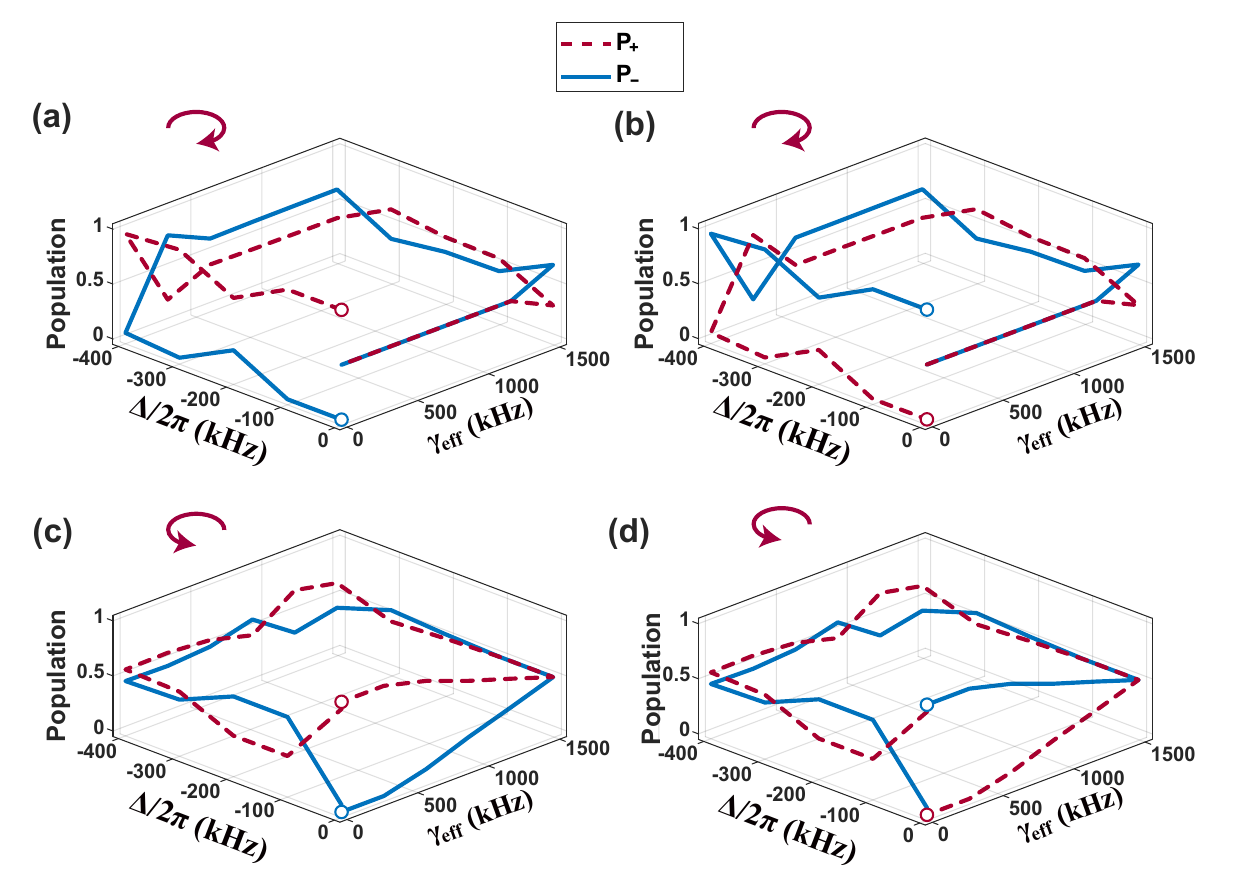}
\caption{Populations $P_{+}$ and $P_{-}$ versus $\Delta$ and $\gamma_{\mathrm{eff}}$ in clockwise and counterclockwise encirclements, where the initial states $|\psi_{+}\rangle$ and $|\psi_{-}\rangle$ are considered respectively, and the typical parameters take the values of $\Delta_{\mathrm{min}}/2\pi=-400$ kHz, $\Delta_{\mathrm{max}}/2\pi=0$ kHz, $\gamma_{\mathrm{min}}\approx0$ kHz, and $\gamma_{\mathrm{max}}\approx1.45$ MHz. The empty circles represent the starting points. }
\label{FigS5}
\end{figure}
\begin{figure}[tbph]
\includegraphics[width=16cm]{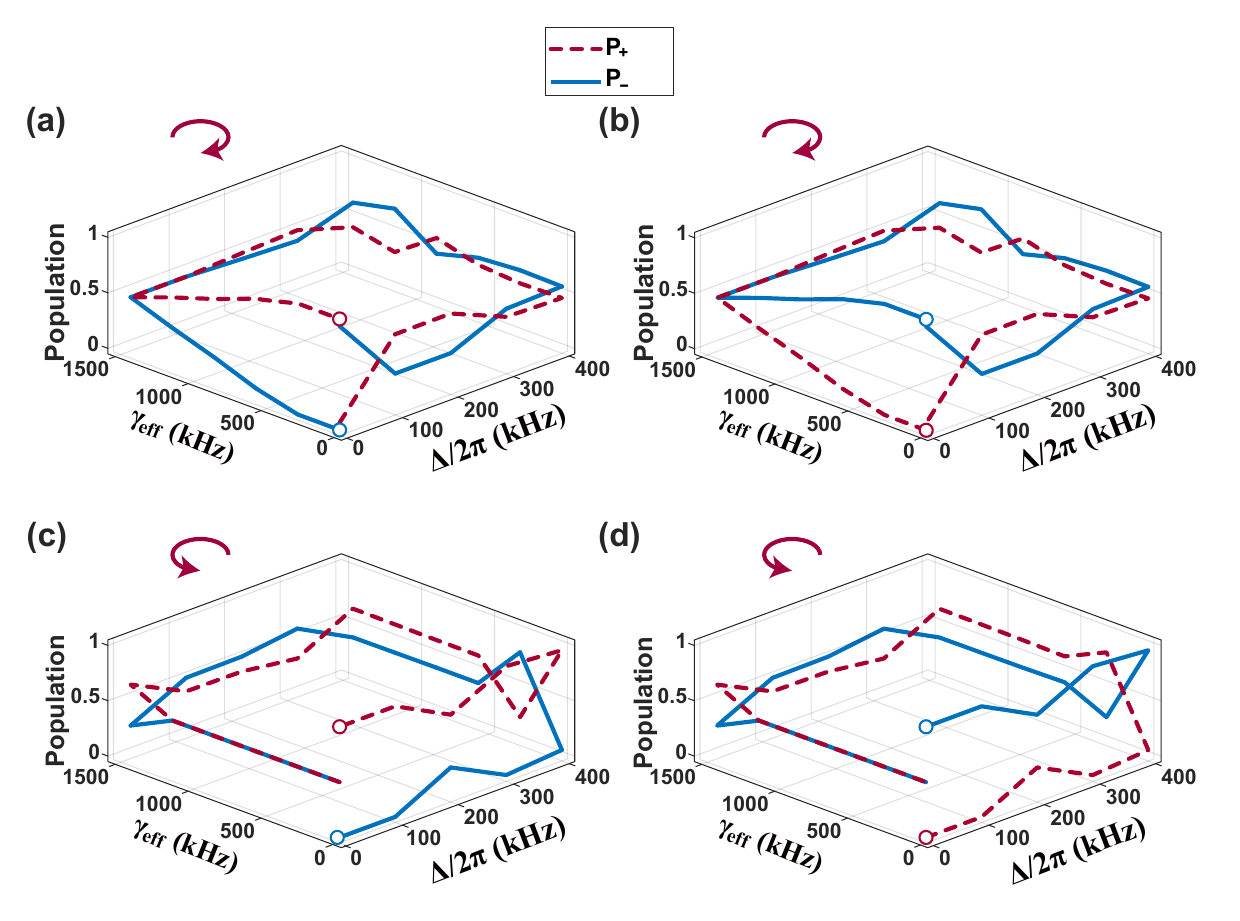}
\caption{Population $P_{+}$ and $P_{-}$  versus $\Delta$ and $\gamma_{\mathrm{eff}}$ in clockwise and counterclockwise encirclements, where the initial states $|\psi_{+}\rangle$ and $|\psi_{-}\rangle$ are considered respectively, and the typical parameters take the values of $\Delta_{\mathrm{min}}/2\pi=0$ kHz, $\Delta_{\mathrm{max}}/2\pi=400$ kHz, $\gamma_{\mathrm{min}}\approx0$ kHz, and $\gamma_{\mathrm{max}}\approx1.45$ MHz. The empty circles represent the starting points. }
\label{FigS6}
\end{figure}

To further understand the relationship between the chirality and the topology, here we consider the encirclement restricted within single Riemann sheets. As such,
we calculate the state evolution in clockwise and counterclockwise encirclements starting from $\Delta=0$ and $\gamma=\gamma_{\mathrm{min}}$ as depicted in Figs.S. \ref{FigS5} and \ref{FigS6}. We choose $t_{d}=6$ $\mu$s and $t_{y}=150$ $\mu$s to keep the change rate of the parameters consistent with our experimental conditions (Here $t_{d}$ and $t_{y}$ represent the durations of iso-decay and isochoric strokes, respectively).

For the clockwise encirclements in Figs.S. \ref{FigS5}(a, b), the initial state $|\psi_{+}\rangle$ or $|\psi_{-}\rangle$ evolves into a steady state that is very similar to the mixture state $(|e\rangle\langle e|+|g\rangle\langle g|)/2$ during the fourth stroke. So neither the closed encirclement nor asymmetric mode convention would appear in this case.
In contrast, for the counterclockwise encirclements starting from either $|\psi_{+}\rangle$ or $|\psi_{-}\rangle$ as depicted by Figs.S \ref{FigS5}(c, d), the system evolves to a steady state at the end of the third stroke, which is very close to the pure state $|g\rangle$ under the condition of large detuning. This steady state experiences a LZS with the detuning increasing from $\Delta/2\pi=-400$ kHz to $\Delta=0$ kHz, reaching $|\psi_{+}\rangle$ finally. Therefore, in this case, starting from $|\psi_{+}\rangle$ yields a closed encirclement, and starting from $|\psi_{-}\rangle$ presents asymmetric mode convention.

For the clockwise encirclements in Figs.S \ref{FigS6}(a, b), the initial state $|\psi_{\pm}\rangle$ evolves into a steady state at the end of the third stroke and finally reaches $|\psi_{-}\rangle$. So  the initial state $|\psi_{-}\rangle$ leads to a closed encirclement, and starting from $|\psi_{+}\rangle$ presents asymmetric mode convention. In contrast, for the counterclockwise encirclements in Figs.S. \ref{FigS6}(c, d), the system from either $|\psi_{+}\rangle$ or $|\psi_{-}\rangle$ finally evolves to a mixed state very similar to $(|e\rangle\langle e|+|g\rangle\langle g|)/2$. So both the closed encirclement and asymmetric mode convention would not appear in this case.

Therefore, when the encirclements are accomplished within single Riemann sheets, the chirality is nonreciprocal and incomplete, which can be called non-reciprocal chirality, i.e., unidirectional chirality.

{\color{black}
\subsection*{VII. More discussions about the chiral behavior and asymmetric mode conversion}
\begin{figure}[tbph]
\includegraphics[width=16cm]{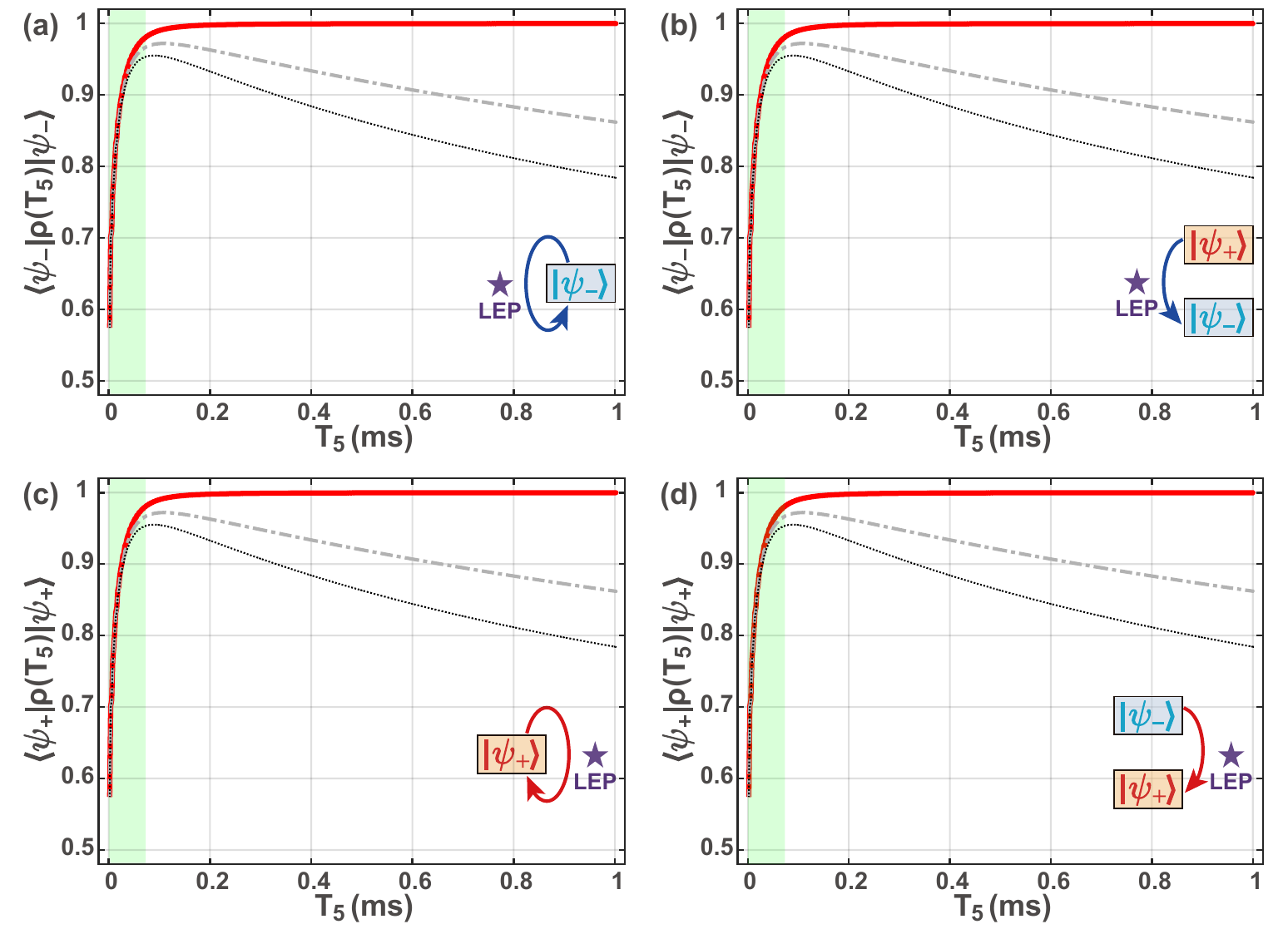}
\caption{Fidelity $\langle\psi_{\pm}|\rho(T_{5})|\psi_{\pm}\rangle$ versus the total evolution time $T_{5}$ in the clockwise and counterclockwise directions, with different initial states $|\psi_{\pm}\rangle$ and quantum jumps. We set $\gamma_{\mathrm{min}}=0$ kHz for the red solid curve, $\gamma_{\mathrm{min}}=25$ kHz for the grey dashed-dotted curve, and $\gamma_{\mathrm{min}}=50$ kHz for the black dotted curve. Other parameters are $\Delta_{\mathrm{min}}/2\pi=1.0$ MHz, $\Delta_{\mathrm{max}}/2\pi=1.0$ MHz, $\gamma_{\mathrm{min}}\approx0$ kHz, and $\gamma_{\mathrm{max}}\approx1.45$ MHz.}
\label{smLEP}
\end{figure}

Now, we numerically explore the impacts of the evolution time and dissipation on chiral behavior and asymmetric mode conversion. After a long-time evolution of the previous four strokes, our system decays to its steady state. Then, the evolution time of the fifth stroke will govern the following evolution and behavior as the Landau-Zener transitions. To clarify the above phenomena, we plot the fidelity $\langle\psi_{\pm}|\rho(T_{5})|\psi_{\pm}\rangle$ versus the total evolution time of the fifth strokes $T_{5}$ as in Fig.S.~\ref{smLEP}.

In the absence of dissipations, the evolution time of the fifth stroke plays a crucial role in determining the adiabaticity. In the non-adiabatic regime, where the evolution time is very short, the system lacks enough time to follow the variation of the detuning. In this situation, non-adiabatic transitions between two eigenstates can occur, leading to variations in the accumulated phase. These phase variations can generate unexpected quantum interference of the Landau-Zener-St{\"u}ckelberg process and weaken the effects of chiral behavior and asymmetric mode conversion (see the shadow areas in green in Fig.S.~\ref{smLEP}).

In contrast, in the adiabatic regime with a long enough evolution time, the system smoothly follows the detuning variation, and the accumulated phase approaches a fixed value. Therefore, adiabatic evolution can suppress non-adiabatic transitions and ensure the manifestation of chiral behavior and asymmetric mode conversion (see the red curves for $\gamma_{\mathrm{eff}}=0$ kHz in Fig.S.~\ref{smLEP}).

However, all experimental results are inevitably associated with dissipations, which lead to quantum jumps. The dissipations in the quantum system lead to a loss of coherence and introduce additional non-adiabatic processes to the ideal Landau-Zener transition. Consequently, dissipations of the system affect the chiral behavior and asymmetric mode conversion. When we set a fixed evolution time, the increased dissipations adjust the chiral behavior towards non-chiral behavior by non-adiabatic transitions (see the gray dashed-dotted cures for $\gamma_{\mathrm{eff}}=25$ kHz and black dotted cures for $\gamma_{\mathrm{eff}}=50$ kHz in Fig.S. \ref{smLEP}).

As a result, our experiments demonstrate chiral behavior and asymmetric mode conversion in general cases, which involve a fixed time and fixed dissipation in the iso-decay strokes. Additionally, the imperfect chiral behavior and asymmetric mode conversion arise from the fact that the conditions for two isochoric strokes cannot follow the conditions for the large detuning.

\begin{figure}[tbph]
\includegraphics[width=16cm]{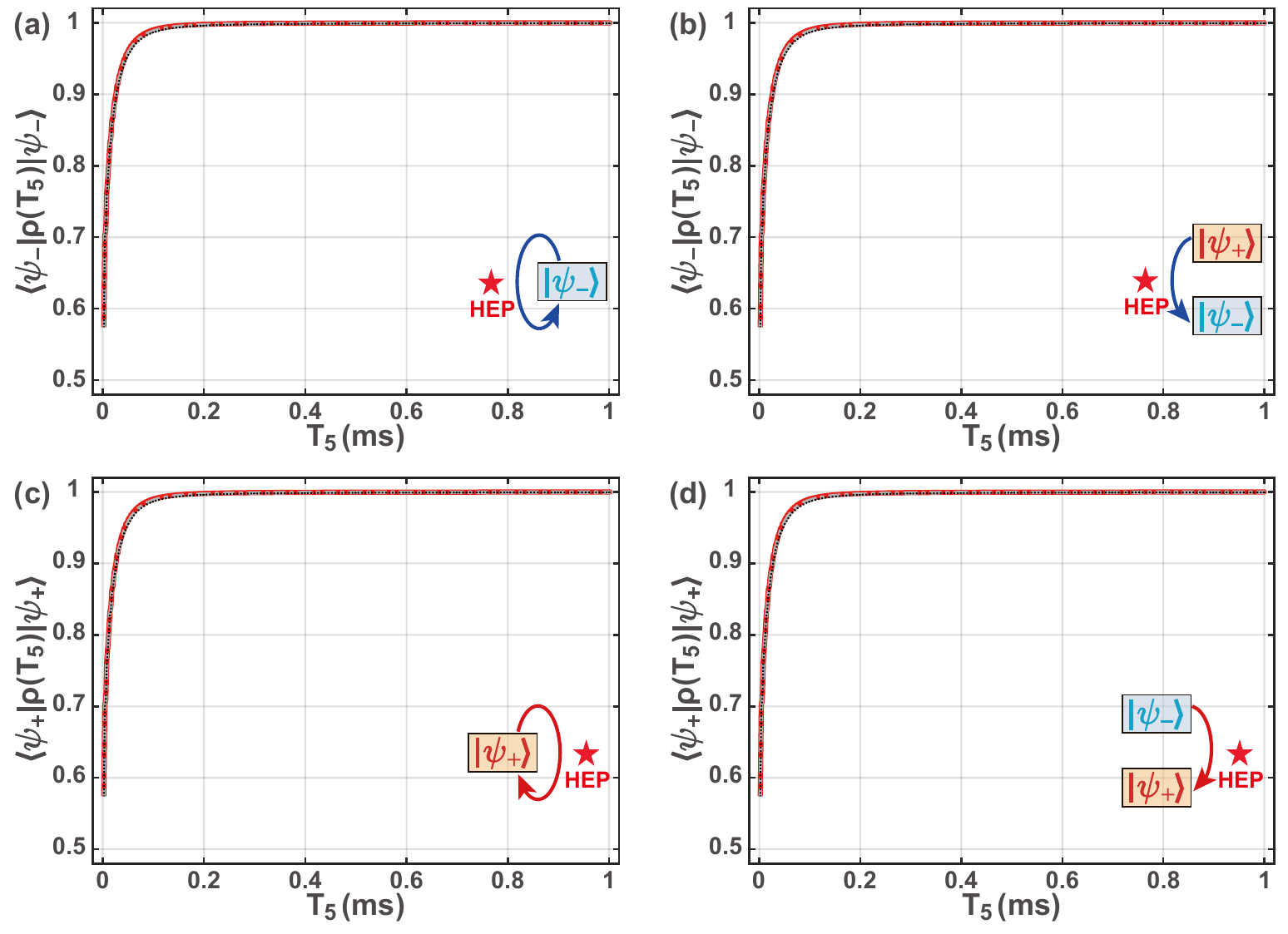}
\caption{Fidelity $\langle\psi_{\pm}|\rho(T_{5})|\psi_{\pm}\rangle$ versus the total evolution time $T_{5}$ in the clockwise and counterclockwise directions, with different initial states $|\psi_{\pm}\rangle$ and no quantum jumps. We set $\gamma_{\mathrm{min}}=0$ kHz for the red solid curve, $\gamma_{\mathrm{min}}=25$ kHz for the grey dashed-dotted curve, and $\gamma_{\mathrm{min}}=50$ kHz for the black dotted curve. Other parameters are $\Delta_{\mathrm{min}}/2\pi=1.0$ MHz, $\Delta_{\mathrm{max}}/2\pi=1.0$ MHz, $\gamma_{\mathrm{min}}\approx0$ kHz, and $\gamma_{\mathrm{max}}\approx1.45$ MHz.}
\label{smHEP}
\end{figure}

On the other hand, with quantum jumps ignored, the above results for LEPs return to those for HEPs. Then, we plot the corresponding results for HEP in Fig.S.~\ref{smHEP}. These figures demonstrate that, without the influence of quantum jumps, we can achieve perfect chiral behavior and asymmetric mode conversion when the evolution time $T_{5}$ is long enough.

These phenomena illustrate that quantum jump is indeed a key quantum feature of the LEP, and our observed chiral behavior and asymmetric mode conversion result from the presence of the LEP.}

\clearpage

\end{document}